\documentclass{article}
\usepackage{amsmath,amssymb,amsfonts}
\usepackage[margin=1in]{geometry}
\usepackage[]{algorithm}
\usepackage[noend]{algorithmic}
\usepackage{graphicx}
\usepackage{lineno}
\usepackage{cite}
\usepackage{xparse}
\usepackage{authblk}
\makeatletter
\let\oldincludegraphics\includegraphics
\RenewDocumentCommand{\includegraphics}{O{} m}{%
  \IfFileExists{#2}{\oldincludegraphics[#1]{#2}}{\fbox{\texttt{\detokenize{#2}}}}%
}
\makeatother
\usepackage{textcomp}
\usepackage{xcolor}
\usepackage{url}
\usepackage{amsmath}
\usepackage{amssymb}
\usepackage{amsthm}
\usepackage{hyperref}
\usepackage[capitalise,noabbrev]{cleveref}
\usepackage{booktabs}
\usepackage[shortlabels]{enumitem}

\usepackage{graphicx}
\usepackage{tikz}
\usepackage{thm-restate}
\usepackage{comment}
\usetikzlibrary{cd}
\nolinenumbers
\newcommand{\set}[1]{\left\{#1\right\}}

\newcommand{\bF}{\mathbb{F}}
\newcommand{\bP}{\mathbb{P}}
\newcommand{\bH}{\mathbb{H}}
\newcommand{\bI}{\mathbb{I}}

\newcommand{\cA}{\mathcal{A}}

\newcommand{\cQ}{\mathcal{Q}}

\newcommand{\ol}{\overline}

\newcommand{\Vmax}{V_{\max}}
\newcommand{\ONDmax}{\OND_{\max}}

\newcommand{\Aquick}{\cA_{\mathrm{quick}}}
\newcommand{\Apf}{\cA_{\mathrm{PF}}}
\newcommand{\Ach}{\cA_{\mathrm{CH}}}

\DeclareMathOperator{\CH}{CH}
\DeclareMathOperator{\ch}{ch}
\DeclareMathOperator{\pf}{pf}
\DeclareMathOperator{\depth}{depth}

\DeclareMathOperator{\quadrant}{quadrant}

\DeclareMathOperator{\CA}{CA}

\DeclareMathOperator{\CF}{CF}
\DeclareMathOperator{\OND}{OND}
\DeclareMathOperator{\OWD}{OWD}
\DeclareMathOperator{\OWND}{OWND}

\DeclareMathOperator{\id}{id}

\DeclareMathOperator{\Universal}{\texttt{Universal}}
\DeclareMathOperator{\UniversalLB}{\texttt{Universal-LB}}
\DeclareMathOperator{\Runtime}{\texttt{Runtime}}

\graphicspath{{figures/}}

\usepackage{todonotes}
\tikzset{notestyleraw/.append style={rectangle}}
 \newtheorem{theorem}{Theorem}
 \newtheorem{lemma}{Lemma}
 \newtheorem{definition}{Definition}
 \newtheorem{corollary}{Corollary}

\def\BibTeX{{\rm B\kern-.05em{\sc i\kern-.025em b}\kern-.08em
    T\kern-.1667em\lower.7ex\hbox{E}\kern-.125emX}}

\title{Tight Adaptive Bounds for Convex Hulls}
\author[1]{Ivor van der Hoog\thanks{\href{https://orcid.org/0009-0006-2624-0231}{ORCID: 0009-0006-2624-0231}}}
\author[1]{Eva Rotenberg\thanks{\href{https://orcid.org/0000-0001-5853-7909}{ORCID: 0000-0001-5853-7909}}}
\author[1]{Daniel Rutschmann\thanks{\href{https://orcid.org/0009-0005-6838-2628}{ORCID: 0009-0005-6838-2628}}}

\affil[1]{IT University of Copenhagen, Denmark}
\date{}
\begin{document}

%\authorrunning{Ivor van der Hoog, Eva Rotenberg, and Daniel Rutschmann}

%\keywords{Convex hull, Combinatorial proofs, Universal optimality}

%\ccsdesc[500]{Theory of computation~Computational geometry}
%\ccsdesc[500]{Theory of computation~Design and analysis of algorithms}

%\funding{This work was supported by the Carlsberg Foundation Fellowship CF21-0302 ``Graph Algorithms with Geometric Applications'', the VILLUM Foundation grant (VIL37507) ``Efficient Recomputations for Changeful Problems'', and the European Union's Horizon 2020 research and innovation programme under the Marie Sk\l{}odowska-Curie grant agreement No 899987. }

%\Copyright{Ivor van der Hoog, Eva Rotenberg, and Daniel Rutschmann}

\maketitle

%\linenumbers
\begin{abstract}
Adaptive sorting algorithms exploit existing order in the input to obtain better-than-worst-case running times. A classical example is sorting by runs: if the input can be partitioned into increasing runs of sizes $s_1, \ldots, s_k$ then the \emph{run-length entropy} is $O(\sum_i s_i \log \frac{n}{s_i})$ and there exist many Merge-sort algorithms which run in this time. 
One can show optimality of such algorithms, by showing that for a fixed sequence of run sizes $s_1, \ldots, s_k$ the worst-case running time of any algorithm lies in $\Omega(\sum_i s_i \log \frac{n}{s_i})$. 

Recently, Eppstein, Goodrich, Illickan, and To introduced algorithms for Pareto fronts, planar convex hulls, and related problems whose running times improve when the input order contains few sorted runs. They analyze the running time algorithm by defining a \emph{Range Partition Entropy} which is a function that depends both the order of the input and the geometric input points. They ask whether matching adaptive lower bounds analogous to those used for run-length entropy can be shown. 

We provide adaptive lower bounds for constructing a convex hull. 
We first observe that if all points lie on the convex hull, a convex hull construction algorithm becomes a sorting algorithm. 
We show that in this case, the algorithm from Eppstein, Goodrich, Illickan, and To is \emph{not} optimal with respect to the run-length entropy. Rather, their algorithm is optimal with respect to its dual: dual run-length entropy.  
We then lift the notion of dual run-length optimality to the plane, and show that their algorithm is optimal under this geometric notion of optimality also. 
Note that  geometric computations can circumvent the $\Omega(n \log n)$ lower bound for their construction for two independent reasons: because the input order is structured (i.e., adaptive to sortedness), or, because the geometry of the instance is simple (e.g., when only few points appear on the Pareto front or convex hull). 
We carefully introduce an optimality measure and lower bounds that are sensitive to both the input's dual run-length entropy and the geometric entropy in a manner which is better than taking the minimum of both. 
\end{abstract}
\thispagestyle{empty}

\setcounter{page}{0}
\newpage
\section{Introduction}
In \emph{adaptive} or \emph{better-than-worst-case} analysis, algorithms are designed to be optimal with respect to structural properties of the input that go beyond just input size. Classical examples include:
\begin{itemize}[noitemsep]
    \item output-sensitive algorithms~\cite{kirkpatrick1986ultimate}, whose running time depends on $n$ and the output size $k$;
    \item structural algorithms~\cite{afshani2009instance, Munro1976Sorting, van_der_hoog_combinatorial_2025}, which fix the input \emph{values} and consider for an algorithm the worst-case running time over all input lists that contain these values in some order;
    \item adaptive sorting algorithms~\cite{auger2019, munro18, juge24, buss19, takaoka2009partial,  schou2024persisort, Castro1992Adaptive, McIlroy1993OptimisticSorting, BarbayNavarro2009CompressedPermutations, BarbayFischerNavarro2012LRMTrees,BarbayNavarro2013CompressingPermutations, GhasemiJugeKhalighinejad2022Galloping, GellingNebelSmithWild2023MultiwayPowersort, BarbayOchoaSatti2017SynergisticMultisets}.
\end{itemize}

Let $P$ be a planar point set of size $n$, and let $I$ be an array storing $P$ in some order. By a standard reduction from sorting, there exists for any convex hull algorithm $\cA$ a point set $P$ and an input order $I$ for which $\cA$ requires $\Omega(n \log n)$ comparisons, which implies a lower bound in any comparison-based model of computation such as the Real RAM. 
For sorting, however, this worst-case bound does not tell the whole story.
The third bullet point above represents a long history of \emph{adaptive sorting analyses}: algorithms that are faster if the input is already somewhat presorted.
For example, if the input array contains many contiguous runs in which the input values are already sorted (e.g., runs of nondecreasing values) then one would expect an algorithm to be faster on such input. 
Adaptive sorting algorithms consider the input array $I$, and choose their favorite aspect of that input array to be adaptive to.
We typically then prove that our algorithms are optimal with respect to the measure they are adaptive to. 
Whilst adaptive analysis is common for sorting algorithms, we believe it has never before been applied to geometric algorithms which reduce to sorting.

\subparagraph{An adaptive convex hull algorithm. }
 For constructing a convex hull (or, Pareto front) there exist linear-time algorithms for when the input array is already sorted by $x$-coordinate~\cite{andrew1979another, Melkman87, graham72}. 
 Eppstein, Goodrich, Illickan, and To~\cite{eppstein_entropy-bounded_2025} asked the natural question of whether a \emph{geometric} algorithm can be made adaptive to how presorted the input is.
 Inspired by Timsort and the run-length entropy that it realizes, they adapted the classical convex hull algorithm by Kirkpatrick, Seidel, and McQueen\footnote{The paper~\cite{kirkpatrick1986ultimate} only lists Kirkpatrick and Seidel as authors. However, in a footnote, McQueen suggested a modification to the algorithm of Kirkpatrick and Seidel that is necessary for the algorithm to become tight. See also the discussion in~\cite{afshani2009instance}.}\cite{kirkpatrick1986ultimate}.
 They provide an upper bound for their algorithm, whose running time is an entropy function that looks similar to the classical run-length entropy from adaptive sorting. They ask whether one can prove, for their algorithm, a notion of adaptive optimality analogous to that established for adaptive sorting algorithms.
 
\subparagraph{Beyond worst-case analysis in geometry.}
Adaptive analysis (analysis adaptive to the presortedness of the input) may be new to computational geometry, but there do exist two prior paradigms in which better-than-worst-case geometric algorithms have been obtained: \emph{output-sensitive} and \emph{structural} analysis. 

In \emph{output-sensitive} analysis, we fix the output size $k$.
For an algorithm $\cA$, its output-sensitive running time $\texttt{Out}(\cA,k)$ is the worst-case running time of $\cA$ over all inputs $I$ whose output has size $k$.
In geometry, there exist output-sensitive analyses for constructing planar convex hulls~\cite{kirkpatrick1986ultimate}, 3D convex hulls~\cite{Chan1996}, 4D convex hulls~\cite{Chan1995Convex}, Voronoi diagrams~\cite{Miller2013newVoronoi}, lower envelopes of monotone chains~\cite{Lu2015}, line segment intersection~\cite{Chazelle1992Optimal}, unions of triangles~\cite{Ezra2005Output} and undoubtedly more. 
For a fixed value $k$, we define the output-sensitive lower bound as: $\min\limits_{\cA} \texttt{Out}(\cA,k)$.
Kirkpatrick and Seidel~\cite{kirkpatrick1986ultimate} showed that when constructing a convex hull, this lower bound is $\Omega(n \log k)$ for every fixed $k$ and the algorithm that they propose matches this lower bound. 

\emph{Structural analysis} is strictly stronger. 
The analysis fixes the set of \emph{values} of the input. It then derives structure from these values and considers for an algorithm it's worst case performance over all inputs with this structure. The classical non-geometric example for structural analysis is \emph{multiset sorting} where the structure are the cliques of duplicates in the input~\cite{Munro1976Sorting}. 
In computational geometry, we are aware of only one prior structural analysis result~\cite{afshani2009instance} and its simplification~\cite{van_der_hoog_combinatorial_2025}.
These papers establish that the algorithm from~\cite{kirkpatrick1986ultimate}, including the footnote by McQueen, is \emph{structurally optimal} (we forward reference Section \ref{sub:results}). This notion of optimality is very technical, but it has a weaker but more intuitive complementary version:
Fix a point set $P$. For any algorithm, it's \emph{order-oblivious} running time is the worst-case running time over all input arrays $I$ that permute $P$. 
The order-oblivious lower bound for any fixed $P$ is the minimum over all algorithms of their order-oblivious running time.
The fact that~\cite{kirkpatrick1986ultimate} is structurally optimal, implies that it is order-oblivious optimal~\cite[Thm. 3.5]{afshani2009instance}: that is, the algorithm (asymptotically) matches the order-oblivious lower bound for every fixed $P$.

 \subparagraph{Contribution.}
In this paper, we formally show adaptive optimality of the convex hull algorithm from~\cite{eppstein_entropy-bounded_2025}.
For many better-than-worst case analyses such as run-length entropy, dual run-length entropy, and structural analysis, there exists a weaker but more intuitive way to phrase the result. 
Since adaptive analysis is a new topic to computational geometry, we will in this introduction state our results in weaker but more intuitive terms. In Section \ref{sub:results}, we give the formal specification of our result. 

Consider first the sorting problem, where the input is an array $I$ storing a set of values.
A more intuitive but weaker version of \emph{run-length} optimality~\cite{Castro1992Adaptive} can be phrased as follows:  
We can define a  \emph{run} in $I$ as a contiguous region of the input where the values appear in sorted order. 
We can partition any input array into runs $R = (s_1, \ldots, s_k)$ and denote by $\Pi_R$ all permutations of $I$ where all regions in $R$ are still runs.
Having fixed $R$ we then provide the weaker but more intuitive definition of the \emph{adaptive running time} of an algorithm $\cA$ as the worst-case running time over all inputs $I' \in \Pi_R$. 
An adaptive algorithm $\cA$ is \emph{optimal} if for every fixed $R$, there exists no algorithm $\cA'$ with a better (asymptotic) adaptive running time.

The convex hull algorithm from~\cite{eppstein_entropy-bounded_2025} is a modification of~\cite{kirkpatrick1986ultimate}.
Intuitively, it runs the recursive algorithm from~\cite{kirkpatrick1986ultimate} where, for each recursive call, it checks in linear time whether the recursive input array is already sorted. If so it runs Graham scan instead of recusing, to terminate in time linear in the current input size. 
This makes the algorithm sensitive to encountering an input that is a \emph{run}, and so the authors from~\cite{eppstein_entropy-bounded_2025} ask whether  a notion of adaptive optimality can be established that is analogous to run-length entropy. 

\begin{figure}[H]
    \centering
    \includegraphics[width = \linewidth]{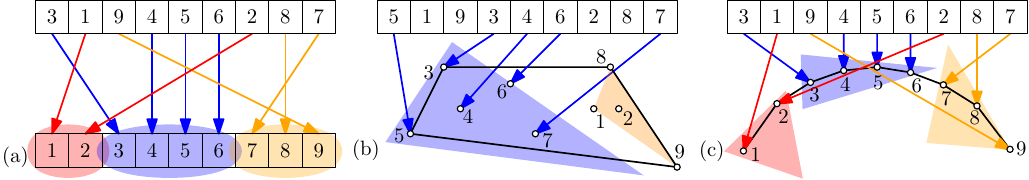}
    \caption{(a) A sorting input array (top) and three dual runs. The yellow run is sorted in decreasing order. 
    (b) Two dual triangles. The yellow triangle is of type (i) and the blue triangle of type (ii), as its points appear in $I$ sorted by $x$. (c) If all points appear on the hull, then type (ii) triangles correspond to dual runs.
    }
    \label{fig:illustration}
\end{figure}

Any convex hull algorithm, ran on inputs where all points lie on the hull, is a sorting algorithm! 
We show that the sorting algorithm induced by~\cite{eppstein_entropy-bounded_2025} is \emph{not} run-length optimal.
Rather, it is dual run-length optimal. This also has a weaker more intuitive version:
Let $I$ be a sorting instance and $O$ be the output. 
We define a \emph{dual run} in $O$ as a contiguous region of the output, where the values in this region appear in  $I$ in sorted order (see Figure~\ref{fig:illustration} (a)).  
We can partition $O$ into dual runs  $R = (s_1, \ldots, s_k)$ and denote by $\Pi_R$ all permutations of $I$ where all regions in $R$ are still dual runs.
One can redefine the \emph{adaptive} running time of an algorithm $\cA$ as the worst-case running time over all inputs $I' \in \Pi_R$. 
An adaptive algorithm $\cA$ is \emph{dual run-length optimal} if for every fixed $R = (s_1, \ldots, s_k)$, there exists no algorithm $\cA'$ with a better (asymptotic) adaptive running time. We prove that the algorithm from~\cite{eppstein_entropy-bounded_2025}, when used as a sorting algorithm, is dual run-length optimal. 

From this notion of dual run-length optimality, we can find a natural geometric analogy. 
However, one has to be careful! If $I$ is an input array storing a planar point set $P$ then not all points in $I$ have to appear in the output. 
Thus, dual runs are tricky to define. One \emph{could} define dual runs using only points on the convex hull, but stronger results are possible. 
Moreover, recall that there are more reasons for why a convex hull algorithm could be faster than $O(n \log n)$: alternatively to sorted runs it may just rely on the fact that $P$ is easy: for example, whenever it has a low order-oblivious lower bound.
The measure that we use must therefore take both $P$ and the order $I$ it arrives in into account.
Our key observation is that the output of a convex hull algorithm is some triangulation where points that lie in the interior of triangles formed by points in $P$ do not appear on the hull. 
Let $I$ be an input array storing a planar point set $P$ (see Figure~\ref{fig:illustration}). We define a dual triangle as any triangle $\Delta$ where either (i)  $\Delta$ is contained in the convex hull or (ii) all points in $P \cap \Delta$ appear in $I$ in sorted order (e.g., by $x$-coordinate). For any input array $I$, we can create a (non-unique) set $R$ of disjoint dual triangles. 
 We can then consider all permutation $\Pi_R$ of $I$ where $R$ is still a set of dual triangles.  We can redefine the \emph{adaptive} running time of an algorithm $\cA$ as the worst-case running time over all inputs $I' \in \Pi_R$.  An adaptive algorithm is then \emph{optimal} if for every fixed set of disjoint triangles $R$, there exists no algorithm $\cA'$ with a better (asymptotic) adaptive running time.  In this paper, we prove that the algorithm from~\cite{eppstein_entropy-bounded_2025} is optimal in this manner (see Section \ref{sub:results} for the stronger statement). 

\newpage
\section{Entropy, Optimality and Theorems}
\label{sec:results}

In this section we state our formal results. 
To this end, we first specify our model of computation.

\subparagraph{Model of computation.}
We use the model of computation from~\cite{van_der_hoog_combinatorial_2025} for our lower bounds. This is a comparison-based model that allows arbitrary comparison-based algorithms, including Real RAM algorithms, but requires every convex hull algorithm to output a \emph{witness}. Such a witness lists the convex hull vertices in cyclic order and, for every non-hull point $p$, gives a triangle spanned by input points whose interior contains $p$.
This differs from~\cite{afshani2009instance}, where the algorithm only has to output the convex hull in cyclic order, but is restricted to algebraic decision trees whose nodes evaluate only constant-complexity bilinear expressions. Both choices have merit; however, requiring a witness makes our lower bounds valid for \emph{any} Real RAM algorithm.

\subparagraph{Entropy.}
The algorithmic analysis will use \emph{Entropy functions}, which are a way to express an algorithm's running time.
Formally, an entropy function takes as input a list $I$ of $n$ values and outputs a number between $0$ and $\log n$.
A classical example is \emph{run-length entropy}~\cite{munro18, juge24, BarbayNavarro2013CompressingPermutations}. 
For the \emph{sorting} problem, the input is a list $I$ of values.
We define an \emph{ordered partition} $\Pi = (I_1, I_2, \ldots)$ as a partition of $I$ into \emph{runs}, where each run $I_i$ is a contiguous sublist of~$I$ whose values are either non-decreasing or non-increasing.

\[
  \text{This induces the \emph{run-length entropy}:} \quad \quad \quad \hfill H_t(I) := 
  \min_{\text{ordered partitions } \Pi \text{ of } I}
  \; \sum_{I_i \in \Pi}   
  \frac{|I_i|}{n}
  \log\frac{n}{|I_i|}. \hfill
\]

Afshani, Barbay, and Chan~\cite{afshani2009instance} study better-than-worst-case performance in a geometric setting.
They revisit the output-sensitive planar convex hull algorithm of Kirkpatrick, Seidel, and McQueen~\cite{kirkpatrick1986ultimate}.
They refine the runtime analysis by defining \emph{structural partitions}.

\begin{definition}
    \label{def:structural_partition}
    Consider an array $I$ containing a point set $P$.
A \emph{structural partition} of $I$ is a list
$\Pi = ((I_1, R_1), (I_2, R_2), \ldots)$
of pairs $(I_i, R_i)$, where $I_i$ is an arbitrary subset of $I$ and $R_i$ is a triangle containing $I_i$ s.t.
(i) $I_1, I_2, \ldots$ partition $I$, and
(ii) each $R_i$ is contained in the closure of the convex hull of $P$.  \vspace{-0.7cm}
\end{definition}

\[
\text{We can then define \emph{structural entropy}~\cite{afshani2009instance}:} \quad \quad \quad  \hfill H_s(I)
  := \min_{\text{structural partitions } \Pi \text{ of } I}
     \sum_{(I_i, R_i) \in \Pi}
     \frac{|I_i|}{n}
     \log \frac{n}{|I_i|}. \hfill
\]

Note that, although we define $H_s(I)$ based on the input list $I$, its value depends only on $P$.
Structural entropy captures the intrinsic difficulty of computing the convex hull of $P$ in a manner that is much more fine-grained than output sensitivity.
Consider, for example, partitioning the unit circle into $\sqrt{n}$ arcs.
For each arc, construct a triangle whose vertices lie on that arc, and place a vertex of $P$ at each corner of the triangle. 
For one of these triangles, place the remaining $\Theta(n)$ points of $P$ arbitrarily inside it.
For any list $I$ containing $P$, the output-sensitive analysis gives the running time $O(n \log n)$ for the algorithm of~\cite{kirkpatrick1986ultimate}, whereas the structural-entropy is much smaller:
$
%\hfill
  H_s(I)
  \leq
  1 \cdot \log 2
  \;+\;
  \sum_{i=1}^{\sqrt{n} - 1}
    \left( \frac{3}{n} \log n \right)
  \;\Rightarrow\;
  O(n(1+H_s(I))) \subseteq O(n). 
  %\hfill
$

Eppstein, Goodrich, Illickan, and To~\cite{eppstein_entropy-bounded_2025}
define an entropy function that combines geometric structural partitioning with the notion of presorted runs in the input. 
Although their entropy definitions differ between the problems that they study, there is a common 
structure.  
We illustrate their entropy function for constructing the planar convex hull which uses \emph{respectful partitions}.

\begin{definition}\label{def:respectful_partition}
    Let $I$ be a list storing a point set $P$.
A \emph{respectful partition} is a list
$\Pi$
of pairs $(I_i, R_i)$, where $I_i$ is a subsequence of $I$ and $R_i$ is a triangle containing $I_i$ such that
(i) the sets $I_1, I_2, \ldots$ form a partition of $I$,
(ii) either $I_i$ is sorted (e.g., by increasing or decreasing $x$-coordinate) or $R_i$ is contained in the convex hull of $P$, and
(iii) if $R_i$ is not contained in the convex hull of $P$ (so $I_i$ is sorted), then $R_i$ is disjoint from all other $R_j$. \vspace{-0.7cm}
\end{definition}

\[
\text{We can then define \emph{range partition entropy}~\cite{eppstein_entropy-bounded_2025}}:  \quad \quad 
\hfill H(I)
:=
\min_{\text{respectful partitions } \Pi \text{ of } I}
\;
\sum_{(I_i, R_i) \in \Pi}
  \frac{|I_i|}{n}
  \log\frac{n}{|I_i|}
. \hfill
\]

\noindent
Eppstein, Goodrich, Illickan, and To~\cite{eppstein_entropy-bounded_2025} design a convex hull algorithm that runs in 
$O(n(1+H(I)))$ time.
Every structural partition is also a respectful partition, and hence for all inputs $I$ we have
$H(I) \leq H_s(I)$. In other words, their algorithm is at least as good as the algorithm from~\cite{kirkpatrick1986ultimate}. 
These two quantities can be separated: e.g., if all points of $P$ lie on the convex hull but $I$ lists all points of $P$ by increasing $x$-coordinate, then $H(I)=0$, and they use $O(n)$ time whilst the classical algorithm~\cite{kirkpatrick1986ultimate} uses $O(n \log n)$ time.

\subsection{Algorithmic optimality}
As adaptive or better-than-worst-case optimality is still relatively new to computational geometry, we first recall how algorithmic optimality is defined. Readers familiar with adaptive sorting or universal optimality may consider skipping to Section~\ref{sub:results}.
An entropy function is, at its core, just a function from inputs to numbers; it is not itself a lower bound. To be algorithmically optimal, an algorithm $\cA$ must have running time asymptotically unbeaten by its peers. Worst-case optimality captures this by considering the worst behavior over all size-$n$ inputs. For bounds beyond the worst case, one must restrict the inputs under consideration, and this restriction is \emph{not} unique.
Haeupler, Wajc, and Zuzic~\cite{Haeupler2021UniversalDistributed} introduced \emph{universal optimality} as an umbrella framework for such better-than-worst-case analyses. It encompasses output-sensitive, run-sensitive, structural, and respectful optimality, and has since gained substantial traction~\cite{Hoog2024Tight, Zuzic2022Universally, Haeupler2021UniversalDistributed, Haeupler2024Dijkstra, Haeupler2025Sorting, hladik2025FORC, Hoog2025SimplerSorting, Hoog2025SimplerDAG, van_der_hoog_combinatorial_2025}. We use this framework to explain our better-than-worst-case bounds.

\subparagraph{The basics.}
Universal optimality considers a family of meaningful universes $U_j$, where each universe $U_j$ is a set of algorithmic inputs of size~$n$.
These universes are not required to be disjoint, but their union must cover all inputs.
For a fixed universe $U_j$, an algorithm~$\cA$ has \emph{universal running time} $
  \texttt{Universal}(\cA, U_j)
  := \max_{I \in U_j} \texttt{Runtime}(\cA, I)$.
Each universe $U_j$ also induces a \emph{universal lower bound} $\texttt{Universal-LB}(U_j) := \min_{\cA'} \texttt{Universal}(\cA', U_j)$.
An algorithm $\cA$ is \emph{optimal} if, for \emph{every} universe $U_j$, the quantity $O(\texttt{Universal}(\cA, U_j))$ matches the corresponding universal lower bound.

The choice for how these universes should be defined is not unique! And to some degree, arbitrary. However the general rule is that as the universes become smaller, it becomes increasingly difficult to design a universally optimal algorithm.
Intuitively, this is because we must construct a \emph{single} algorithm that, for every universe $U_j$, competes against an algorithm that may be tailor-made for~$U_j$.

One can derive information-theoretic lower bounds for a fixed universe $U_j$ as follows.
Let $O_j$ denote the set of distinct outputs produced on inputs $I \in U_j$.
Any algorithm $\cA$ operating on $U_j$ induces a decision tree whose leaves correspond to these outputs.
Hence, there exists at least one input $I \in U_j$ for which the running time of $\cA$ is in $\Omega\bigl(\log |O_j|\bigr)$
since it must distinguish among $|O_j|$ possible outputs.
Thus, the information-theoretic lower bound for $U_j$ shows $\texttt{Universal-LB}(R) \in \Omega\bigl(\log |O_j|\bigr)$.
This is the basic tool that we deploy, what remains is to argue that we have a novel and more useful notion of universe. 

\subparagraph{Universal optimality for sorting.}
In the introduction, we explained that many adaptive sorting algorithms have a more intuitive but weaker formulation. Here, we provide the technical but stronger version using universal optimality. 
As a warm-up, consider \emph{multiset sorting}. Let $I$ store a multiset. We can partition $I$ into maximal parts of equal values, $\Pi = (I_1, I_2, \ldots)$. The \emph{multiset entropy} is
$
H_m(I)
:= \sum_{I_i \in \Pi}
\frac{|I_i|}{n}
\log\!\frac{n}{|I_i|}.
$
The multiset sorting algorithm of Munro~\cite{Munro1976Sorting} runs on every input $I$ in time $O(n(1+ H_m(I)))$.
This running time can be interpreted through universal optimality by letting the universe depend on input multiplicities. For a size sequence $(s_1, \ldots, s_k)$, define the corresponding universe as all inputs $I$ whose multiset partition satisfies $|I_i| = s_i$ for every $i$. For each fixed sequence $(s_1, \ldots, s_k)$, output counting gives the universal lower bound
$
\Omega\!\left(
n + \sum_{i=1}^k s_i \log\!\frac{n}{s_i}
\right).
$
Thus, Munro's algorithm is universally optimal under this choice of universes. However, there are incomparable choices of universes, and even strictly stronger ones.

Next, we present the stronger and more technical form of run-length optimality. Let $(s_1,\ldots,s_k)$ be a sequence of sizes summing to $n$, and define its universe as all inputs $I$ for which each block
$
I[s_1+\cdots+s_{\ell-1} : s_1+\cdots+s_{\ell}]
$
is sorted. For this fixed universe, output counting gives the lower bound
$
\Omega!\left(n + \sum_{i=1}^k s_i \log\frac{n}{s_i}\right).
$
For an input $I$ in this universe, $O(n(1+H_t(I)))$ can be much smaller than this bound. However, it is never asymptotically larger and so any algorithm running in $O(n(1+H_t(I)))$ time is universally optimal. This universe is incomparable to the multiset universe: optimality for one does not imply optimality for the other.

Our final example is the stronger, more technical form of dual run-length optimality~\cite{GhasemiJugeKhalighinejad2022Galloping}. Again let $(s_1,\ldots,s_k)$ be a sequence of sizes summing to $n$. Define its universe as all inputs $I$ such that the first $s_1$ elements of the final sorted order appear increasingly in $I$, the next $s_2$ elements appear increasingly in $I$, and so on. Output counting gives the universal lower bound
$
\Omega!\left(n + \sum_{i=1}^k s_i \log\frac{n}{s_i}\right),
$
so any algorithm running in $O(n(1+H_s(I)))$ time is universally optimal. Since every set of duplicates induces a dual run, $H_s(I) \le H_m(I)$, making this universe strictly stronger than the multiset universe.

\subsection{Defining our problems, entropy, universes, and results}
\label{sub:results}

We analyze three algorithms from~\cite{eppstein_entropy-bounded_2025}: its induced sorting algorithm, Pareto front algorithm, and convex hull algorithm. For each algorithmic problem, we define a suitable universe to show two things:
(i) universal optimality and (ii) that the entropy function $H(I)$ from~\cite{eppstein_entropy-bounded_2025} is squeezed in-between our universal upper and lower bound.
Note that (ii) is not necessary for optimality, but it does show that the authors from~\cite{eppstein_entropy-bounded_2025} deployed the right tool to analyze their algorithms. Due to its lesser importance, (ii) is deferred to Section~\ref{sec:entropy}.

\subparagraph{Sorting.}
Let the input be a list $I$ of $n$ values.
We  observe that we can equivalently define the dual run-length entropy  $H(I)$ from~\cite{GhasemiJugeKhalighinejad2022Galloping}  by projecting $I$ onto the unit circle to obtain $I'$, and then returning the range partition entropy of $I'$ (Definition~\ref{def:respectful_partition}). 
From there, it is folklore that truncated Quicksort (i.e., what~\cite{eppstein_entropy-bounded_2025} does if all points lie on the hull) is dual run-length optimal. 
As a warm-up, we illustrate in Section~\ref{sec:sorting}:

\begin{restatable}{theorem}{quicksortthm}
\label{thm:quicksort}
Fix a number sequence $(s_1,\ldots,s_k)$ that sums to $n$ and define the universe $\mathbb{I}_{s_1,\ldots,s_k}$ as all lists $I$ of length $n$ where the first $s_1$ elements of the final sorted order appear increasingly in $I$, the next $s_2$ elements appear increasingly in $I$, etc. 
  (i) the sorting algorithm $\cA$ from~\cite{eppstein_entropy-bounded_2025} is optimal, i.e., $\texttt{Universal}(\cA,(s_1,\ldots,s_k))$ $
\;\le\; 
c \cdot
\texttt{Universal-LB}(s_1,\ldots,s_k)$. 
(ii) $\texttt{Universal-LB}(s_1,\ldots,s_k) \in \displaystyle\max_{I \in \mathbb{I}_{s_1, \ldots, s_k}} \Omega( n(1+H(I)) )$.
\end{restatable}

\subparagraph{Pareto front.}
Next, we study the Pareto front construction algorithm from~\cite{eppstein_entropy-bounded_2025}. Whilst our ultimate goal is to analyze convex hulls, Pareto fronts offer a geometrically simpler alternative. 
The input is a list $I$ storing a point set $P$ in general position.
The \emph{Pareto front} consists of points not dominated by others in linear order. 
We require as output the Pareto front and a \emph{witness list} which contains for each index $i$ where $I[i]$ is not on the Pareto front an index $j$ such that $I[j]$ dominates $I[i]$. 

The paper~\cite{eppstein_entropy-bounded_2025} overloads the definitions of respectful partitions and entropy, and we will match their overloaded terminology. Since we use Pareto fronts as our simpler example, we will restrict our analysis to considering only \emph{increasing} sorted sequences.
That means that in the context of Pareto fronts,
a \emph{respectful partition} is a list $\Pi=((I_1,R_1),\ldots)$ where each $I_i$ is a subsequence of $I$, and each $R_i$ is an axis-aligned rectangle that contains $I_i$, such that
(i) the $I_i$ form a partition of $I$,
(ii) either $I_i$ is sorted in increasing order (e.g.\ by $x$), or, the top-right corner of $R_i$ is dominated by a point of~$P$, and
(iii) if the top-right corner of $R_i$ is not dominated, then $R_i$ is disjoint from all other $R_j$.
\[
\text{This yields the entropy~\cite{eppstein_entropy-bounded_2025}:} \quad \quad \quad \hfill
H(I)
:=
\min_{\text{respectful }\Pi}
\sum\nolimits_{(I_i,R_i)\in\Pi}
\frac{|I_i|}{n}\log\!\frac{n}{|I_i|}.
\hfill
\]
\noindent
To show optimality, we define a suitable universe.
For the Pareto front, we define a universe $\bI_{P, R}$ by a point set $P$ and disjoint axis-aligned rectangles $R$ that cover $P$; it contains all inputs $I$ storing $P$ such that, for each $r\in R$, the subsequence $I\cap r$ is monotone in $x$ or $y$ (Definition~\ref{def:pareto_universe}). 
We show:

\begin{restatable}{theorem}{paretoTHM}
\label{thm:pareto}
Fix a point set $P$ and list of ranges $R= (R_1, \ldots, R_k)$ and denote by $\mathbb{I}_{P, R}$ the resulting universe.
  Then (i) the Pareto front algorithm $\cA$ from~\cite{eppstein_entropy-bounded_2025} is optimal, i.e., $\texttt{Universal}(\cA,(P, R))
\;\le\; 
c \cdot
\texttt{Universal-LB}(P, R)$. 
Moreover (ii) $\texttt{Universal-LB}(P, R) \in \displaystyle\max_{I \in \mathbb{I}_{P, R}} \Omega( n(1+H(I)) )$.
\end{restatable}
\vspace{-0.3cm}

\subparagraph{Convex hull.}
The main algorithmic problem in this paper is the construction of a planar convex hull. The input is again a list $I$ storing $P$.
The output lists the convex hull and a witness list containing for each index $i$ where $I[i]$ is not on the convex hull three indices $(a, b, c)$ such that $I[i]$ is contained in the triangle $(I[a], I[b], I[c])$.
We define a \emph{respectful partition} as a list $\Pi=((I_1,R_1),\ldots)$ according to Definition~\ref{def:respectful_partition}. 

\[
\text{Recall that this yields the \emph{range partition entropy}~\cite{eppstein_entropy-bounded_2025}: } \quad \quad \quad
\hfill
H(I)
:=
\min_{\text{respectful }\Pi}
\sum\nolimits_{(I_i,R_i)\in\Pi}
\frac{|I_i|}{n}\log\!\frac{n}{|I_i|}.
\hfill
\]
To show optimality, we define a suitable universe $\mathbb{I}_{P, R}$ by a point set $P$ and disjoint triangles $R$ that cover $P$; it contains all inputs $I$ storing $P$ such that, for each $r\in R$, the subsequence $I\cap r$ is monotone in $x$ or $y$.

\begin{restatable}{theorem}{convexTHM}
\label{thm:convex}
Fix a point set $P$ and list of triangles $R= (R_1, \ldots, R_k)$ that cover $P$ and denote by $\mathbb{I}_{P, R}$ the resulting universe.
  Then (i) the convex hull  algorithm $\cA$ from~\cite{eppstein_entropy-bounded_2025} is optimal, i.e., $\texttt{Universal}(\cA,(P, R))
\;\le\; 
c \cdot
\texttt{Universal-LB}(P, R)$. 
Moreover (ii) $\texttt{Universal-LB}(P, R) \in \displaystyle\max_{I \in \mathbb{I}_{P, R}} \Omega( n(1+H(I)) )$.
\end{restatable}

\subparagraph{Implications.}
Every respectful partition (Definition~\ref{def:respectful_partition}) is also structural (Definition~\ref{def:structural_partition}), but the converse is not true. Hence, by Theorem~\ref{thm:convex} the convex hull algorithm is at least as strong as the structurally optimal or order-oblivious algorithms of~\cite{afshani2009instance}. Additionally, it is then adaptive to the presortedness of the input array.

\subsection{Sorting}
\label{sec:sorting}

If we apply the Pareto front or convex hull algorithms from~\cite{eppstein_entropy-bounded_2025} to inputs that reduce from sorting, it results in the \emph{truncated quicksort} depicted in \cref{algo:quicksort}. 

\begin{algorithm}[h] 
\caption{truncated-quicksort$(I')$.}\label{algo:quicksort}
\begin{algorithmic}[1]
    \REQUIRE $I' \text{ is a sublist of } I$.
    \IF{$I'$ is sorted in increasing order}
        \STATE \textbf{return} $I'$
    \ELSIF {$I'$ is sorted in decreasing order }
        \STATE \textbf{return} $\operatorname{reversed}(I')$
    \ENDIF
    \STATE $m \gets \operatorname{median} I'$
    \STATE $I_1, I_2 \gets \operatorname{stable-partition}(I', m)$
    \STATE \textbf{return} $\text{truncated-quicksort}(I_1) + m + \text{truncated-quicksort}(I_2)$
\end{algorithmic}
\end{algorithm}

\begin{lemma} \label{lemm:quicksort_upper}
    Fix $(s_1, \ldots s_k)$ and input $I$ in this universe. 
  Algorithm~\ref{algo:quicksort} runs in time $O(n(1 + \sum_{i} \frac{s_i}{n} \log \frac{n}{s_i} ))$.
\end{lemma}
\begin{proof}
    The time spent in one call to truncated-quicksort$(I')$, excluding recursive calls, is $O(|I'|)$.
    Let $M$ be the total size of all sublists $I'$ across all such calls,
    then total running time of truncated quicksort is $O(M)$.
    Let $i \in \{1, \dots, k\}$ be arbitrary and let $A_i$ be the set of items
    of rank $[s_1 + \dots + s_{i-1}, s_1 + \dots + s_{i})$ in $I$.
    Since $I$ is from the universe $(s_1, \ldots s_k)$, these items form an increasing or decreasing subsequence of $I$.
    At recursion depth $d$, there are at most four calls to truncated-quicksort for which $I'$ contains some elements of $A_i$.
    (All except at most two of these calls terminate early due to $I'$ being sorted.)
    Thus, at most $\min(s_i, 4 n / 2^{d})$ elements of $A_i$ participate in the $d$-th recursion level.
    \begin{align*}
        M &\le \sum_{i=1}^{k} \sum_{d=1}^{\lfloor\log n\rfloor} \min(s_i, 4 n / 2^{d}) \le \sum_{i=1}^{k} \Big(\sum_{d=1}^{\lfloor\log \frac{n}{s_i}\rfloor}s_i  + \sum_{d=\lfloor \log \frac{n}{s_i} \rfloor+1}^{\infty} \frac{4 n}{2^d}\Big)
        %&\le \sum_{i=1}^{k} \Big(s_i \log \frac{n}{s_i} + 8 s_i \Big) = 
        \leq n\Big(8 + \sum_{i=1}^{k} \frac{s_i}{n} \log \frac{n}{s_i}\Big). \qedhere
    \end{align*}
\end{proof}

\begin{lemma} \label{lemm:quicksort_lower}
     For a universe $(s_1, \ldots s_k)$, $\texttt{Universal-LB}((s_1, \ldots s_k)) \in \Omega(n(1 + \sum_{i} \frac{s_i}{n} \log \frac{n}{s_i} ))$.
\end{lemma}
\begin{proof}
    Any algorithm needs $\Omega(n)$ time to read the input.
    Denote by $D$  the number of distinct outputs over all inputs $I$ in the universe $(s_1, \ldots s_k)$.
    Then $\log D$ is an information theoretic lower bound and
    \[
      \texttt{Universal-LB}((s_1, \ldots s_k)) \ge \log D \geq \log \frac{n!}{\prod_{i=1}^{k} s_i!} =  \sum_{i} s_i \log \frac{n}{s_i}  - O(n).
    \]
    Combining this equation with the $\Omega(n)$ lower bound proves the lemma. 
\end{proof}
Combining \cref{lemm:quicksort_upper,lemm:quicksort_lower} proves Theorem~\ref{thm:quicksort} (i). For (ii), we refer to Section~\ref{sec:entropy}.

\quicksortthm*

\newpage
\section{Pareto Front} \label{sect:pareto}

We denote by $\bP_n$ all sets of $n$ planar points in general position,
that is, no two points have the same x- or y-coordinate and no three points lie on a line.
We denote by $\pf(P)$ the set of points that lie on the Pareto front of $P$ in their linear order.  $\bI_P$ denotes all arrays of length $n$ that contain $P$ in arbitrary order.

We consider algorithms that construct the Pareto front of $I$ and require that the algorithm to output two lists: The \emph{front list}
encodes $\pf(P)$, and the \emph{witness list} encodes a witness for every point in $P - \pf(P)$.
More precisely, a \emph{front list} $F = (f_1, \dots, f_k)$ is a sequence of integers such that
$I[f_1], \dots, I[f_k]$ are precisely the points in $\pf(P)$, ordered by increasing x-coordinate.
A \emph{witness list} $W = (w_1, \dots, w_n)$ is a sequence of $n$ integers such that $w_i = -1$ if $I[i] \in \pf(P)$,
and $I[w_i]$ dominates $I[i]$ otherwise.

\subparagraph{Sorted Regions.} 
The authors from~\cite{eppstein_entropy-bounded_2025} introduce \emph{sorted regions}.
For computing the Pareto front, they define a \emph{region set} $R$ to be a set
of disjoint axis-aligned rectangles in the plane. 

\begin{definition}
    \label{def:pareto_universe}
    Let $P$ be a point set. An input $I \in \bI_P$ \emph{respects $R$} if
for each region $r \in R$, the points in $r \cap P$ when ordered along their appearance in $I$ are sorted by x-coordinate or y-coordinate.
We define a universe $(P, R)$ by fixing $P$ and a region set $R$ that covers $P$, and we define by $\bI_{P, R}$ all inputs $I \in \bI_P$ that respect $R$.
\end{definition}

We note that for our analysis, we restrict the above definition to sequences that are sorted by $x$ or $y$ in increasing order. Our convex hull result from Section~\ref{sect:hull} is more involved and alsos allow decreasing sequences.
 Given Definition~\ref{def:pareto_universe} and universe $(P, R)$, the \emph{universal running time} is defined as
$
    \Universal(\cA, (P, R)) = \max\limits_{I \in \bI_{P, R}} \Runtime(\cA, I).$
An algorithm $\cA$ is \emph{universally optimal} 
if there is a constant $c>0$ such that, for all universes $(P, R)$, and all algorithms $\cA'$,
we have $\Universal(\cA, P, R) \le c \cdot \Universal(\cA', P, R)$.

\subsection{Lower Bounds}

If an algorithm $\cA$, across some family of inputs,
produces $k$ distinct outputs, then the maximum running time of $\cA$ over this family of inputs is $\ge \log k$.
In our setting, the output consists of both a front list and a witness list.
We will apply the information theoretic lower bound to each of these separately.

Fix a universe $(P, R)$. 
Observe that, for every input $I \in \bI_{P, R}$, there is one front list.
We denote by $\bF_{P, R}$ the set of all front lists across all inputs $I \in \bI_{P, R}$, then the information theoretic lower bound is $\log | \bF_{P, R}|$.
Witness lists are more complicated.
There can be multiple witness lists that are valid for a single input.
Our lower bound counts the number of \emph{inputs} for which a given list is a witness list: 

\begin{definition}
    Fix a universe $(P, R)$ and a list $W \in \mathbb{Z}^n$ of $n$ integers. We define:
    \[
        V(P, R, W) = \set{I \in \bI_{P, R} \mid W \text{ is a witness list for } I} \text{ and }
        \Vmax(P, R) = \max_{W \in \mathbb{Z}^n} |V(P, R, W)|.
    \]
\end{definition}
\begin{theorem} \label{theo:pareto_information_theoretical}
    Fix a universe $(P, R)$. Then, for any algorithm $\cA$,
    \[
        \Universal(\cA, (P, R)) \in \Omega\Big(n + \log |\bF_{P, R}| + \log \frac{|\bI_{P, R}|}{\Vmax(P, R)}\Big).
    \]
\end{theorem}
\begin{proof}
    Any algorithm needs $\Omega(n)$ time to read the input, so $\Universal(\cA, (P, R)) \in \Omega(n)$.
    Since every input $I \in \bI_{P, R}$ has one front list,
    the algorithm $\cA$ needs to output each different front list in $|\bF_{P, R}|$  at least once
    over all inputs in $\bI_{P, R}$.
    The information theoretic lower bound shows $\Universal(\cA, (P, R)) \in \Omega(\log |\bF_{P, R}|)$.
    Finally, we consider the witnesses.
    A list of integers $W \in \mathbb{Z}^n$ is a witness for at most $\Vmax(P, R)$ different inputs in $\bI_{P, R}$.
    Thus, the algorithm $\cA$ needs to output at least $\lceil \frac{|\bI_{P, R}|}{\Vmax(P, R)}\rceil$
    different lists $W$ across all inputs in $\bI_{P, R}$.
    The information theoretic lower bound shows $\Universal(\cA, (P, R)) \in \Omega(\log \frac{|\bI_{P, R}|}{\Vmax(P, R)})$.
    Since $n$, $\log |\bF_{P, R}|$, and $\frac{|\bI_{P, R}|}{\Vmax(P, R)}$ are non-negative and $\Universal(\cA, (P, R))$ is asymptotically bigger than each of these individual values, the theorem follows. 
\end{proof}

\subsection{Ordered Normalized Downdrafts}

We estimate the quantity $\frac{|\bI_{P, R}|}{\Vmax(P, R)}$ through combinatorial objects that we call ordered normalized downdrafts. 
These are technical, but intuitively capture for a fixed universe $(P, R)$ all ways a $p \in P - \pf(P)$ could elect its witness in a manner that is sensitive to $R$.
Let $P$ be a point set:

\begin{definition}
    \label{def:downdraft}
    A \emph{downdraft}~\cite{van_der_hoog_combinatorial_2025} is a map $\varphi : P - \pf(P) \to P$
such that for each $p \in P - \pf(P)$, the point $\varphi(p)$ dominates $p$.
\end{definition}

\begin{definition} \label{def:compass}
    Let $R$ be a region set.
    A \emph{compass function} $\kappa$ for $R$ specifies, for each $r \in R$, how the points in $I \cap r$ should be sorted.
    Formally, it is a map
$
        \kappa : R \to 
        \{\text{``increasing by $x$''},\;
          \text{``increasing by $y$''} \}.
    $
\end{definition}

We write $\CF(P, R)$ for the set of all compass functions on $R$ given the point set $P$.

\begin{lemma} \label{lemm:pareto_linear_order}
    Let $P$ be a point set, $R$ a region set, and $\kappa \in \CF(P, R)$.
    We can construct a linear order $L$ on $P$ such that:
    \begin{itemize}[noitemsep]
        \item If $p,q\in P$ and $p$ dominates $q$, then $q <_L p$.
        \item If $r\in R$ is sorted by $x$ under $\kappa$ and $p,q\in P\cap r$ with $p.x < q.x$, then $p <_L q$.
        \item If $r\in R$ is sorted by $y$ under $\kappa$ and $p,q\in P\cap r$ with $p.y < q.y$, then $p <_L q$.
    \end{itemize}
\end{lemma}

\begin{figure}
    \centering
    \includegraphics[]{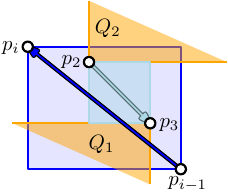}
    \caption{The final step in the proof of \cref{lemm:pareto_linear_order}.} \label{fig:red_blue}
\end{figure}

\begin{proof}
Let $G$ be the directed graph on the vertex set $P$ with edges defined as follows:
\begin{itemize}[noitemsep]
    \item For any $p,q\in P$ with $p$ dominating $q$, add a \emph{red} edge $q \to p$.
    \item If $r\in R$ is $x$-sorted under $\kappa$ and $p,q\in P\cap r$ with $p.x<q.x$, add a \emph{light blue} edge $p \to q$.
    \item If $r\in R$ is $y$-sorted under $\kappa$ and $p,q\in P\cap r$ with $p.y<q.y$, add a \emph{dark blue} edge $p \to q$.
\end{itemize}

    An edge is \emph{blue} if it is light blue or dark blue.
    If, for any $p, q \in P$, there is both a red edge $p \to q$ and a blue edge $p \to q$, then we delete the blue edge
    and only keep the red edge.
    If $G$ is a DAG, then an arbitrary topological ordering of $G$ yields the linear order $L$.
    What remains is to show that $G$ is acyclic.

    Suppose for the sake of contradiction that $G$ is not acyclic and  let $C = p_1 p_2 \dots p_k p_1$ be a shortest undirected cycle in $G$. 
    If both $p_i p_{i+1}$ and $p_{i+1} p_{i+2}$ are red edges, then so is $p_i p_{i+2}$,
    contradicting the minimality of $C$.
    Similarly, if both $p_i p_{i+1}$ and $p_{i+1} p_{i+2}$ are blue edges,
    then these edges are both light blue or both dark blue, depending on which region $p_{i+1}$ lies in (recall that regions are disjoint), but then $p_i, p_{i+2}$ is an edge (possibly red), contradicting the minimality of $C$.
    We may hence assume that $k$ is even and that $p_1 p_2, p_3 p_4, \dots$ are red edges,
    and $p_2 p_3, p_4 p_5, \dots$ are blue edges.
    By minimality of $C$, these are the only edges between vertices on $C$.
    Without loss of generality, suppose that $p_2 p_3$ is a light blue edge. See \cref{fig:red_blue}.
    Since $p_3$ has no incoming red edges, there are no points of $C$ in the bottom-left quadrant $Q_1$ centered at $p_3$.
    Similarly, since $p_2$ has no outgoing red edges, there are no points of $C$ in the top-right quadrant $Q_2$ centered at $p_2$.
    On $C$, let $p_i$ be the next point after $p_3$ with $p_i.x < p_3.x$. Then $p_{i-1}.x > p_3.x$, so $p_{i-1}p_i$ is a dark blue edge.
    But then, the region in $R$ that contains $p_2 p_3$ and the region in $R$ that contains $p_{i-1} p_i$
    intersect.
    This contradiction shows that $G$ is a DAG, so a linear extension $L$ exists.
\end{proof}

\subparagraph{Normalized  ordered downdrafts.} Let $P$ be a point set, $R$ a region set, $\kappa$ a compass function for $R$,
and $L$ the linear order from \cref{lemm:pareto_linear_order}.

\begin{definition}
    \label{def:ordered_normalized_downdraft}
A \emph{normalized downdraft} is a downdraft (see Definition~\ref{def:downdraft}) $\psi$ such that for all regions $r \in R$ and
points $p, q \in (P - \pf(P)) \cap r$ with $p <_L q$, we have $\psi(p) \leq_L \psi(q)$.
\end{definition}

Given any normalized downdraft $\psi$ and point $q \in P$, the \emph{fiber} $\psi^{-1}(\{q\})$ is the set of all points $p \in P - \pf(P)$ that get mapped to $q$.
Given the linear order $L$ from \cref{lemm:pareto_linear_order}, we define the \emph{ordered normalized downdraft} $\ol{\psi} = (\psi, \prec_{\ol{\psi}})$  as a normalized downdraft $\psi$ plus 
a total order $\prec_{\ol{\psi}}$ on each fiber $\psi^{-1}(\{p\})$,
such that for all regions $r \in R$ and points $p, q \in (P - \pf(P)) \cap r$ with $\psi(p) = \psi(q)$,
we have $p <_L q \Leftrightarrow p \prec_{\ol{\psi}} q$.
We denote by $\OND(P, R, \kappa)$ the set of all ordered normalized downdrafts and $\ONDmax(P, R) = \max\limits_{\kappa \in \CF(P, R)} |\OND(P, R, \kappa)|$.
In the remainder of this section, we upper-bound the running time of~\cite{eppstein_entropy-bounded_2025} in terms of $\ONDmax(P,R)$.
In Section~\ref{sub:counting}, we relate this quantity to $\frac{|\bI_{P,R}|}{\Vmax(P,R)}$ to establish tightness.

\subsection{The Eppstein-Goodrich-Illickan-To Algorithm for Pareto Front}
\label{sub:counting}

The Pareto front algorithm from~\cite{eppstein_entropy-bounded_2025} is a recursive algorithm that is defined as in Algorithm~\ref{algo:pareto_front}.

\begin{algorithm}[h] 
\caption{Eppstein-Goodrich-Illickan-To$(I')$.}\label{algo:pareto_front}
\begin{algorithmic}[1]
%    \REQUIRE $I' \text{ is a sublist of } I$.
    \IF{$I'$ is sorted}
        \STATE compute $\pf(I')$ and a witness for each point not on $\pf(I')$ in linear time
        \STATE \textbf{return} $\pf(I')$ to the front list
    \ENDIF
    \STATE $m \gets \operatorname{median} \{p.x \mid p \in I'\}$
    \STATE $q \gets \arg \max \{p.y \mid p \in I': p.x \ge m\}$ \hfill ($q$ is the representative)
    \STATE $I_1 \gets \{p \in I' \mid p.y > q.y\}$
    \STATE $I_2 \gets \{p \in I' \mid p.x > q.x\}$
    \STATE assign to all points in $I'$ that are dominated by $q$ the witness $q$. 
    \STATE Eppstein-Goodrich-Illickan-To$(I_1)$
    \STATE add $q$ to the front list
    \STATE Eppstein-Goodrich-Illickan-To$(I_2)$
\end{algorithmic}
\end{algorithm}

\begin{figure}[h]
    \centering
    \includegraphics[]{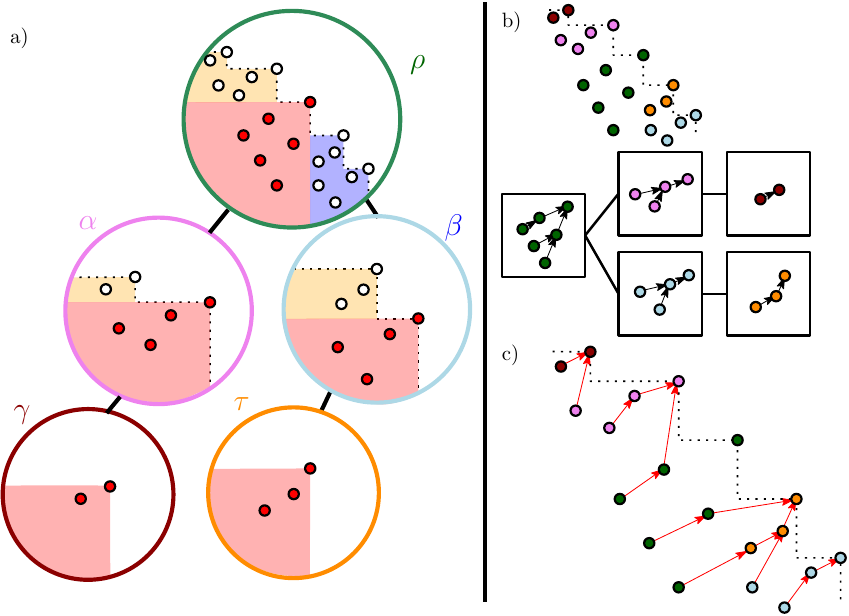}
    \caption{(a) A point set $P$ with its quadrant tree.
(b) We color each $p \in P$ by $u(p)$.
(c) For $R=\emptyset$, the normalized downdraft $\psi$ maps every $p \in P - \pf(P)$ to a point that dominates it.
% \vspace{-0.9cm}
    }
    \label{fig:quadrant_tree}
\end{figure}

\noindent
We analyze the running time of Algorithm~\ref{algo:pareto_front} through a \emph{quadrant tree} (see Figure~\ref{fig:quadrant_tree}). 

\begin{definition}
    Let $P$ be a point set.
    A \emph{quadrant tree}  of $P$ is a binary tree where the root stores some $p \in \pf(P)$.
    Its \emph{quadrant} are all points dominated by $p$, and its \emph{population} is $P \cap \quadrant(p)$.
    The left child is the quadrant tree of all points in $P$ whose $y$-coordinate is strictly larger than that of $p$.
    The right child is the quadrant tree of all points in $P$ whose $x$-coordinate is strictly larger than that of $p$.
 Every $p \in P$ is in the population of exactly one node, which we denote by $u(p)$. 
 Given a set $R$ of rectangles, the \emph{truncated quadrant tree} moreover makes a node a leaf if all remaining points $P'$ are contained in a single region $r \in R$. The population of such a leaf are all points in $P'$, and there is no representative.
\end{definition}

The execution of \cref{algo:pareto_front} depends on the order in which $P$ appears in $I$.
Let $R$ be a region set.
We modify \cref{algo:pareto_front} by replacing the condition ``$I'$ is sorted'' with ``$P'$ lies entirely in some region $r \in R$''.
The resulting algorithm depends only on $P, R$.
Its recursion tree is a truncated quadrant tree $Q(P,R)$, whose nodes correspond to recursive calls.
% We assign to each quadrant tree node the representative point $q$ of its corresponding call.

\begin{lemma} \label{lemm:pf_running_time}
Fix a universe $(P,R)$ and let $I \in \bI_{P,R}$.
Let $Q(P,R)$ be the truncated quadrant tree induced by the modified \cref{algo:pareto_front}.
Then the original \cref{algo:pareto_front} on $I$ runs in time
$O(n + \sum\limits_{p\in P} \depth_{Q(P,R)}(u(p)))$,
where $\depth_{Q(P,R)}(u(p))$ is the depth of $u(p)$ in $Q(P,R)$.
\end{lemma}

\begin{proof}
    We compare the running time of \cref{algo:pareto_front} with the running time of the modified algorithm.
    If $P'$ is contained in a single region, then $I'$ is always sorted.
    Thus, the running time of \cref{algo:pareto_front} is upper-bounded by the running time of the modified algorithm.
    Each recursion call of the modified algorithm spends constant time per point in $P'$.
    Thus, the running time of the modified algorithm is bounded by $
        O\Big(n + \sum\limits_{p \in P} \left( \depth_{Q(P, R)}  ( u(p) ) \right)\Big).$
\end{proof}

\begin{figure}
\centering
    \includegraphics[]{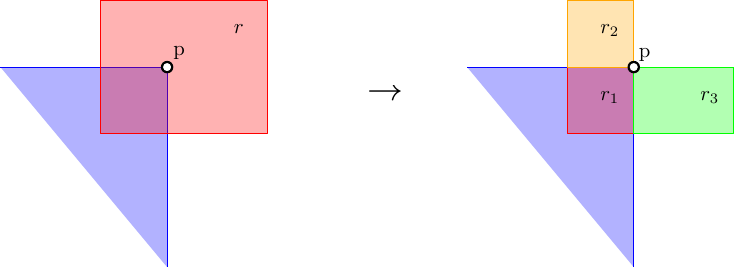}
    \caption{Cutting a region $r$ into three pieces.} \label{fig:refinement}
\end{figure}

\newpage
A region $r \in R$ might intersect the population of many nodes in the quadrant tree.
We will define a region set $R'$ such that every region in $r'$ intersects the population of at most one node in $Q(P, R)$:
For every region $r \in R$, start at the root node $u$ of $Q(P, R)$.
Remove from $r$ the interior of the area that dominates the representative of $u$.
Then split $r$ into at most three pieces by defining $r_1$ as the intersection of $r$ with the quadrant of $u$ and by $(r_2, r_3)$ the remaining areas (\cref{fig:refinement}).
Recursively apply this procedure to obtain $R'$.

\begin{lemma} \label{lemm:quadrant_tree_ond}
    Fix the universe $(P, R)$ and let $R'$ be the region set derived from $R$ by the above procedure.  
    Consider the truncated quadrant tree $Q(P, R')$ induced by running the modified Algorithm~\ref{algo:pareto_front} on $(P, R')$. \vspace{-0.3cm}
    \[
     \hfill  \text{  Then}  \quad \quad \quad \hfill   \ONDmax(P, R') \le \frac{n^n}{2^{\sum_{p \in P} \left( \depth_{Q(P, R')}  ( u(p) )\right)}} \cdot \prod_{r \in R'} \frac{1}{|r \cap P|!} \hfill
    \]
\end{lemma}

\begin{proof}
    Fix a compass function $\kappa \in \CF(P, R')$.
    Let $L'$ be the linear order from \cref{lemm:pareto_linear_order} for the triple $(P, R', \kappa)$.
    We say a \emph{weak downdraft} is a map $\tau : P - \pf(P) \to P$
    such that, for every point $p \in P - \pf(P)$,
    the point $\tau(p)$ lies in the subtree of $u(p)$.
    An \emph{ordered weak downdraft} is a weak downdraft $\tau$ plus a linear ordering $\le_{\ol{\tau}}$ on each fiber. 
    A \emph{weak normalized downdraft} is a weak downdraft $\tau$ where 
    for $r \in R'$ and $p, q \in P \cap r$ with $p <_L q$,
    we have $\tau(p) \neq_L \tau(q)$ (we refer to this as Condition A).
    An \emph{ordered weak normalized downdraft} is a weak normalized downdraft plus a linear order on each fiber,
    such that for all $r \in R'$ and points $p, q \in P \cap r$ with $\tau(p) = \tau(q)$,
    we have $p <_L q \Leftrightarrow p \prec_{\ol{\tau}} q$ (we call this Condition B).

    We denote by $\OWD(P, R')$ the set of all ordered weak downdrafts,
    and by $\OWND(P, R', \kappa)$ the set of all ordered weak normalized downdrafts.
    Every ordered normalized downdraft is a  ordered weak normalized downdraft. Thus, we can upper bound the number of ordered normalized downdrafts $\OND(P, R', \kappa)$ by  $ |\OND(P, R', \kappa)| \le |\OWND(P, R', \kappa)|$.
    Let $r \in R'$ be a region and $\tau \in \OWD(P, R', \kappa)$.
    For any permutation $\sigma \in S_{r \cap P}$, the map $\tau \circ \sigma$ is also
    a weak ordered downdraft. Indeed, all points $p \in r \cap P$ have the same $u(p)$,
    so they place identical restrictions on the weak downdraft. (This would not be true for ordered downdrafts.) 
    Moreover, there is exactly one permutation $\sigma \in S_{r \cap P}$
    such that the resulting map $\tau \circ \sigma$ satisfies the conditions (A) and (B)
    for this region $r$. 
    
    This argument applies to every region $r \in R'$, thus
    $
        |\OWND(P, R', \kappa)| =  \prod\limits_{r \in R'} \frac{1}{|r \cap P|!} \cdot |\OWD(P, R')|. 
    $
 
    \[
       \text{   What remains is to show:} \quad \OWD(P, R') \le \frac{n^n}{2^{\sum_{p \in P} \left( \depth_{Q(P, R')} \cdot u(p) \right)}}
    \]
    Consider the truncated quadrant tree $Q(P, R')$.
    Every ordered weak downdraft can be constructed bottom up.
    For a node $u \in Q(P, R')$, we first construct an ordered weak downdraft $\tau_1$ from the left child of $u$,
    and an ordered weak downdraft $\tau_2$ from the right child of $u$.
    Then, to define an ordered weak downdraft for all points in the subtree of $u$,
    we start from $\tau_1 \sqcup \tau_2$ and place each point $p$ in the population of $u$ into some fiber.
    For any such point $p$, there are most $m(u)$ options where $m(u)$ is the number of points in the subtree of $u$.
    Since \cref{algo:pareto_front} partitions at the median, $m(u) \le n / 2^{\depth(u)}$. This shows the lemma.
\end{proof}

\begin{theorem} \label{theo:pareto_upper_bound}
    The running time of \cref{algo:pareto_front} is $
        O\Big(n + \log \frac{|\bI_{P, R}|}{\ONDmax(P, R)} \Big).$
\end{theorem}

\begin{proof}
    By \cref{lemm:pf_running_time}, the running time is bounded by  $
        O\Big(n + \sum\limits_{p \in P} \depth_{Q(P, R)} \left( u(p) \right) \Big)$.
        
    We observe that since $R'$ is a refinement of $R$,  it follows that for any $p \in P$, we have $\depth_{Q(P, R)} (u(p)) \leq \depth_{Q(P, R')} (u(p))$, and so we can apply \cref{lemm:quadrant_tree_ond} and obtain:
    \[
        \sum_{p \in P} \depth_{Q(P, R)} (u(p))  \le \sum_{p \in P} \depth_{Q(P, R')} (u(p))   \leq \log \Big(\frac{n^n}{\ONDmax(P, R')} \cdot \prod_{r \in R'} \frac{1}{|r \cap P|!} \Big).
    \]
    Since $R'$ is a refinement of $R$, we furthermore have $\ONDmax(P, R') \ge \ONDmax(P, R)$.
    We apply an algebraic argument identical to the last step of \cref{lemm:quicksort_upper} to derive the following upper bound:
    
    \[
        \prod_{r \in R'} \frac{1}{|r \cap P|!} \le 2^{O(n)} \cdot \prod_{r \in R} \frac{1}{|r \cap P|!}. \hfill \text{ By Stirling's approximation, $n^n \le n! \cdot 2^{O(n)}$.
    Therefore:}
    \]

    \[
        \sum_{p \in P} \depth_{Q(P, R)} u(p) \le \log \Big(\frac{1}{\ONDmax(P, R)} \cdot \frac{n!}{\prod_{r \in R} |r \cap P|!}\Big) + O(n)
    \]
    Finally, $\bI_{P, R}$ contains all inputs for which the points in every region are sorted by increasing x-coordinate, thus
    $
        \frac{n!}{\prod\limits_{r \in R} |r \cap P|!} \le |\bI_{P, R}|
    $ which shows the lemma. 
\end{proof}

\noindent
What remains is to show that $
        O\Big(n + \log \frac{|\bI_{P, R}|}{\ONDmax(P, R)} \Big)$ matches $\Omega\Big(n + \log |\bF_{P, R}| + \log \frac{|\bI_{P, R}|}{\Vmax(P, R)}\Big)$.

For any fix pair $(P, R)$, we have four quantities that we have defined:
first $|\bI_{P, R}|$, the number of lists $I$ that store $P$ for which $R$ respects $I$  (Definition~\ref{def:pareto_universe}).
Second $\ONDmax(P, R)$, the maximum number of ordered normalized downdrafts induced by a single compass function.
Third $|\bF_{P, R}|$, the number of front lists across all inputs $I \in \bI_{P, R}$.
Finally $\Vmax(P, R)$, the maximum number of inputs $I \in \bI_{P, R}$ that can share a witness list. 
We relate these quantities to each other by chasing a \emph{commuting diagram}. We warn the reader: this argument (and its more involved counterpart for convex hulls) is the most technical part of this paper and we have found no way to reduce its technicality.

\newpage
\subsection{Showing that our upper bound matches the lower bound.}

\begin{theorem}
    \label{thm:pareto_map}
    Fix the universe $(P, R)$ and list $W \in \mathbb{Z}^n$. 
    Recall that  $V(P, R, W)$ denotes the inputs $I \in \bI_{P, R}$ for which $W$  is a witness list for $I$.     There is a natural map
    \[
        \Phi : V(P, R, W) \hookrightarrow \bigcup_{\kappa \in \CF(P, R)} \Big(\{\kappa\} \times \OND(P, R, \kappa) \times \bF_{P, R}\Big), \hfill \text{    and this map is injective.}
    \]

\end{theorem}

\begin{proof}
    Let $I \in V(P, R, W)$. We view $I$ as a map $[n] \mapsto P$ and we construct $\Phi$:
    
    \paragraph{Defining the map $\Phi$.} There is a unique front list $f \in \bF_{P, R}$ for $I$.
    We define a compass function $\kappa$ as follows: for each $r \in R$,
    $\kappa(r)$ describes the order in $I \cap r$.
    e.g. whether the points in $P \cap r$ appear in $I$ in order sorted by $x$-coordinate. If $p \cap R$ is sorted increasing by both $x$ and $y$, $\kappa(r)$ chooses the sorting by $x$.

    We define a downdraft $\varphi$ as follows (see Definition~\ref{def:downdraft}): For every integer $i$, if $W[i] \ne -1$, put $\varphi(I[i]) = I[W[i]]$.
    Since $W$ is a witness, this defines a downdraft.
    Next, we define a permutation $\sigma$ that permutes the points inside each region,
    and use $\sigma$ to turn $\varphi$ into a normalized downdraft $\psi$.
    Let $L$ be the linear order from \cref{lemm:pareto_linear_order}.
    For any region $r$, let $p_1 <_L \dots <_L p_k$ be the points in $r \cap (P - \pf(P))$ sorted by $L$,
    and let $q_1, \dots, q_k$ be the same set of points, but sorted by $\varphi(q_1) \le_L \dots \le_L \varphi(q_k)$.
    If there are ties, say $\varphi(q_i) = \varphi(q_{i+1})$ for some index $i$, then we require that $q_i <_L q_{i+1}$.
    We put $\sigma(p_i) = q_i$ and $\psi = \varphi \circ \sigma$.

    Finally, we use $\psi$ to define an ordered normalized downdraft $\ol{\psi} = (\psi, <_{\ol{\psi}})$ by ordering each fiber of $\psi$ (see Definition~\ref{def:ordered_normalized_downdraft} and the paragraph below it):
    Let $i, j \in [n]$ with $\varphi(I[i]) = \varphi(I[j])$.
    Then $\psi(\sigma^{-1}(I[i])) = \varphi(I[i]) = \varphi(I[j]) = \psi(\sigma^{-1}(I[j]))$,
    so $\sigma^{-1}(I[i])$ and $\sigma^{-1}(I[j])$ lie in the same fiber of $\psi$.
    We put $\sigma^{-1}(I[i]) \prec_{\ol{\psi}} \sigma^{-1}(I[j])$ if and only if $i < j$.
    Put $\Phi(I) = (\kappa, \ol{\psi}, f)$. By doing this for every $I \in V(P, R, W)$, we get for each input $I$ a compass function, an ordered normalized downdraft, and the unique front list $f$ which defines the map $\Phi$.

    \paragraph{Showing a contradiction.}
        For any $I \in V(P, R, W)$, our definition of $\Phi(I)$ creates some pair $(\sigma, \varphi)$ which leads to the output of $\Phi(I)$. 
    Similarly, any $I' \in V(P, R, W)$ creates some pair $(\sigma', \varphi)$ which leads to $\Phi(I')$.
    Suppose for the sake of contradiction that $\Phi$ is not injective. Then for two distinct $I, I'  \in V(P, R, W)$, $\Phi(I) = \Phi(I') = (\kappa, \ol{\psi}, f)$. 
    We immediately note that then these two inputs $I$ and $I'$ `share' $\kappa$, $\ol{\psi}$, the front list $f$, the point set $P$, witness list $W$, set of regions $R$ that cover $P$, \emph{and} the linear order $L$ constructed by the proof of Lemma~\ref{lemm:pareto_linear_order}. 
      Note that  \emph{a priori}, we do not know whether $\sigma = \sigma'$ or $\varphi = \varphi'$.   However, we \emph{can} conclude the following fact from the construction of $\Phi$. For any index $i \in [n] - f$, we have:
    \begin{equation}
        \psi(\sigma^{-1}(I[i])) = \varphi(I[i]) =  I[W[i]] \quad \quad \text{ and } \quad \quad
    \psi(\sigma^{-1}(I'[i])) = \varphi'(I'[i]) =  I'[W[i]]
    \end{equation}

    Since there are distinct integers $i, i' \in [n]$ with $I[i] = I'[i'] = u$, we pick the pair $(i, i')$ so that $u$ is maximal with respect to $<_L$.
    That is, there exists no distinct integers $i_0, i_0' \in [n]$ with $I[i_0] = v = I'[i'_0]$ with $u <_L v$.     
    We first consider the case where $u \in \pf(P)$, say, $u$ is the $k$'th point on $\pf(P)$.
    Since $\Phi(I) = \Phi(I')  = (\kappa, \ol{\psi}, f)$, they both share the same front list $f$. But then $I[f[k]] = u = I'[f[k]]$ which implies $i' = f[k] = i$, a contradiction.
    Therefore, $u \notin \pf(P)$.

    \paragraph{Creating a commuting diagram.}
  The remainder of our proof relies on a commuting diagram. 
  We define the shared functions $r_W$ and $r_\psi$, and a function   $r_{\varphi}$ and $r_{\varphi'}$ for $I$ and $I'$ respectively (see Figure~\ref{fig:commuting_diagram}). 
 
    Both inputs have the same witness list $W$. Define $r_W$ as the map that assigns every $i \in [n] - f$ its rank in the fiber $W^{-1}(\{W[i]\})$.
    I.e., let $W[i] =j$. Then since $W$ is a witness list of $I$ and $I'$, it is true that $I[i]$ is dominated by $I[j]$ and $I'[i]$ is dominated by $I'[j]$. 
    Moreover, the index $j$ has a fixed set of indices $W^{-1}(\{ W[i] \} ) = W^{-1}(\{ j\}) = \{ i, ... \}$ 
    where for each $k \in W^{-1}(\{ j\})$, $I[k]$ is dominated by $I[j]$ and $I'[k]$ is dominated by $I'[j]$. 
    The function $r_W$ considers the set of indices $W^{-1}(\{ W[i] \})$ sorted from low to high and returns the rank of $i$ in this sorted order. 
    Similarly, let $r_{\psi}$ be the map that assign every $u \in P - \pf(P)$ its rank
    in the fiber $\psi^{-1}(\{\psi(u)\})$ when the fiber is sorted by $\prec_{\ol{\psi}}$.
    Analogously, we define $r_{\varphi}$ and $r_{\varphi'}$ by ordering each fiber by its position along $I$.
    By the $\prec_{\ol{\psi}}$-part of the definition of $\Phi$,
    for any $i \in [n] - f$, we have $r_\psi(\sigma^{-1}(I[i])) = r_{\varphi}(I[i]) = r_W(i)$.
    Similarly, $r_\psi(\sigma^{-1}(I'[i])) = r_{\varphi'}(I'[i]) =  r_W(i)$.
    Thus, the diagram in \cref{fig:commuting_diagram} commutes.

\begin{figure}[h]
\centering
    \includegraphics[]{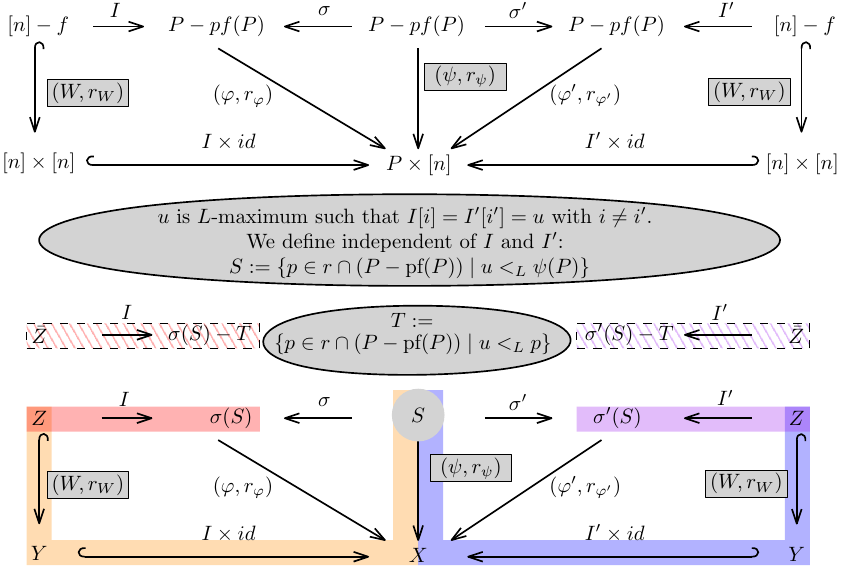}
    \caption{Top: the commuting diagram. Grey elements are `shared' between the distinct inputs $I, I' \in V(P, R, W)$ based on the fact that $\Phi(I) = \Phi(I')$. Bottom: The diagram chasing, starting from $S$.\vspace{-0.5cm}} \label{fig:commuting_diagram}
\end{figure}

    \paragraph{Chasing our commuting diagram.}
    If $u \not \in \pf(P)$ then there is a unique region $r \in R$ that contains $u$. 
    Let $S = \set{p \in r \cap (P - \pf(P)) \mid u <_L \psi(p)}$.
    We now chase the diagram, starting with the orange path in \cref{fig:commuting_diagram}.
    Starting from $S$, we may apply $(\psi, r_\psi)$ to get a set $X \subseteq P \times [n]$,
    then apply $(I \times \id)^{-1}$ to get a set $Y \subseteq [n] \times [n]$, then apply $(W, r_W)^{-1}$
    to get a set $Z \subseteq [n] - f$. Since the diagram commutes and the maps $(I \times \id)$ and $(W, r_W)$ are injective,
    we have $\sigma^{-1}(I(Z)) = S$.

We can alternatively apply the blue path in Figure~\ref{fig:commuting_diagram}, starting from $S$, applying the shared pair $(\psi, r_\psi)$ to obtain $X$. We then apply $(I' \times \id)^{-1}$,
and then $(W,r_W)^{-1}$.
For every $v \in \psi(S)$ we have $u <_L v$ by definition of $S$.
Recall that $u$ was chosen as the maximal point along $L$ for which there were distinct integers $i,i'\in[n]$ with $I[i]=u=I'[i']$. Therefore, every point $v$ with $u<_L v$ appears at the same index in $I$ and $I'$ and this is true in particular for all $v \in \psi(S)$. 
Hence $(I\times\id)^{-1}(X)=(I'\times\id)^{-1}(X)$, and so both
routes through the diagram yield the same sets $Y$ and $Z$.

  By the definition of $\Phi$, we have
$\varphi(\sigma(p))=\psi(p)$ and $r_\varphi(\sigma(p))=r_\psi(p)$.
Since $W$ is the witness list for $I$, we can chase the red path in the diagram to map $p\in S$ to the index of $\sigma(p)$ in $I$. Hence $I[Z]=\sigma(S)$.
Analogously, the purple path maps each $p\in S$ to the index of
$\sigma'(p)$ in $I'$, and hence $I'[Z]=\sigma'(S)$.

    We define the set $T := \{p \in r \cap (P - \pf(P)) \mid u <_L p\}$.
    Note that the definition of $T$ is independent of $I$ and $I'$.
    We have $\sigma(S) = \{p \in r \cap (P - \pf(P)) \mid u <_L \varphi(p)\}$ by definition of $S$ and the fact that $\psi = \varphi \circ \sigma$. 
For every $p \in T \cup \{u\}$, the point $\varphi(p)$ dominates $p$, and hence
$p <_L \varphi(p)$. Since $u \leq_L p$, it follows that
$u <_L \varphi(p)$, so $p \in \sigma(S)$. Therefore,
$T \cup \{u\} \subseteq \sigma(S)$. Similarly,
$T \cup \{u\} \subseteq \sigma'(S)$.
    Recall that $u$ was chosen as the maximal point along $L$ for which there were distinct integers $i,i'\in[n]$ with $I[i]=u=I'[i']$. 
    Therefore, $I^{-1}(T) = I'^{-1}(T)$.
    Put $\ol{Z} = Z - I^{-1}(T)$, then applying the red path and purple paths again implies that $I[\ol{Z}] = \sigma(S) - T$ and $I'[\ol{Z}] = \sigma'(S) - T$. 
    Observe that $u$ is the $<_L$-maximal point in $\sigma(S) - T$ and in $\sigma'(S) - T$,
    and that $\sigma(S) - T \subseteq r$ and $\sigma'(S) - T \subseteq r$.
    Hence, if the compass map $\kappa$ sorts the points in $P \cap r$ in increasing order (by $x$- or $y$-coordinate), $i = \max(\ol{Z}) = i'$.
    If $\kappa$ sorts in decreasing order, then $i = \min(\ol{Z}) = i'$.
    In either case, $i = i'$.
    This contradiction shows that $\Phi$ is injective.
\end{proof}

\begin{corollary} \label{coro:pareto_counting}
    Let $P$ be a point set and $R$ a region set.
    Then
    \[
        \Vmax(P, R) \le |\bF_{P, R}| \cdot \ONDmax(P, R) \cdot 4^{n}.
    \]
\end{corollary}
\begin{proof}
    Regions $r \in R$ that contain no points of $P$ have no effect on $\Vmax(P, R)$ or $\OND(P, R)$.
    Thus, we may assume that every region in $R$ contains a least one point in $P$,
    so $|R| \le n$.
    By Theorem~\ref{thm:pareto_map}, for every witness $W$
    \[
        V(P, R, W) \le \sum_{\kappa \in \CF(P, R, W)} |\OND(P, R, \kappa)| \cdot |\bF_{P, R}| \le |\CF(P, R, W)| \cdot \ONDmax(P, R) \cdot |\bF_{P, R}|.
    \]
    Since $|\CF(P, R, W)| \le 4^{|R|} \le 4^n$,
    the corollary follows by taking the maximum over $W$.
\end{proof}

\paretoTHM*

\begin{proof}[Proof of Theorem~\ref{thm:pareto} (i)]
    By \cref{theo:pareto_upper_bound}, the universal running time of \cref{algo:pareto_front} is $
        O\Big(n + \log \frac{|\bI_{P, R}|}{\ONDmax(P, R)}\Big)$.
    By \cref{coro:pareto_counting}, $\displaystyle\quad \ONDmax(P, R) \ge \frac{\Vmax(P, R)}{|\bF_{P, R}| \cdot  4^{n}}$, so the running time of \cref{algo:pareto_front} is
    $\displaystyle O\Big(n + \log \frac{|\bI_{P, R}|}{\Vmax(P, R)} + \log |\bF_{P, R}|\Big)$, which matches the lower bound from \cref{theo:pareto_information_theoretical}.\qedhere
\end{proof}

\section{Convex Hull} \label{sect:hull}

Our main result is a universal lower and upper bound for constructing a planar convex hull.
The proof follows along similar lines of the proof in Section~\ref{sect:pareto}, but is significantly more involved. This is in part because witnesses are more complicated for convex hulls (as they are no longer a singular point), and also our downdrafts are much more intricate.
This section uses the same compass functions as Definition~\ref{def:compass} but in Section~\ref{sec:generalization} we show how to generalize our result to allow for other types of sorted orders (where in the most general form, we only require that for a region $r \in R$ the points in $P \cap r$ form a simple polygonal curve when ordered along the input $I$). 
Recall that, for any integer $n$, $\bP_n$ denotes all sets of $n$ points in the plane that lie in general position.
For a point set $P$, the convex hull is the smallest convex region that contains $P$.
We denote by $\ch(P)$ the set of points in $P$ that lie on the convex hull, in cyclic order.
Recall that $\bI_P$ denotes all arrays of length $n$ that contain $P$ in some order.
 Recall that we define a suitable universe $\mathbb{I}_{P, R}$ by a point set $P$ and disjoint triangles $R$ that cover $P$; it contains all inputs $I$ storing $P$ such that, for each $r\in R$, the subsequence $I\cap r$ is monotone in $x$ or $y$. Universal running times fix $(P, R)$ and take the worst-case running time of an algorithm over all inputs $I \in \mathbb{I}_{P, R}$.

\subparagraph{The Output.} We again require the algorithm to output two lists: The \emph{hull list} encodes $\ch(P)$,
and the \emph{witness list} encodes a witness for every point in $P - \ch(P)$.
More precisely, a \emph{hull list} $H = (h_1, \dots, h_k)$ is a sequence of integers such that $I[h_1], \dots, I[h_k]$ are the points in $\ch(P)$, in cyclic order starting with the point with minimal $x$-coordinate. A \emph{witness list} $W = ((a_1, b_1, c_1), \dots, (a_n, b_n, c_n))$ is a sequence of $3n$ integers such that $a_i = b_i = c_i = -1$ if $I[i] \in \ch(P)$,
and otherwise, $I[i]$ lies inside the triangle $\Delta(I[a_i], I[b_i], I[c_i])$.

\subparagraph{Universal Lower Bounds.} Our approach follows a similar approach to Section~\ref{sect:pareto} where we provide a universal lower bound by counting how many inputs fit a specific witness list. Fix a universe $(P, R)$.
For every input $I \in \bI_{P, R}$, there is one hull list. We define
\[
    V(P, R, W) = \set{I \in \bI_{P, R} \mid W \text{ is a witness list for } I} \text{ and }
    \Vmax(P, R) = \max_{W \in \mathbb{Z}^{n \times 3}} |V(P, R, W)|.
\]
\begin{theorem}
    Fix a universe $(P, R)$. Then, for any convex hull algorithm $\cA$,
    \[
        \Universal(\cA, (P, R)) \in \Omega\Big(n + \log |\bH_{P, R}| + \log \frac{|\bI_{P, R}|}{\Vmax(P, R)}\Big).
    \]
\end{theorem}
\begin{proof}
    Any algorithm needs $\Omega(n)$ time to read the input, so $\Universal(\cA, (P, R)) \in \Omega(n)$.
    Since every input $I \in \bI_{P, R}$ has one hull list,
    the algorithm $\cA$ needs to output each different hull list in $|\bH_{P, R}|$  at least once
    over all inputs in $\bI_{P, R}$.
    The information theoretic lower bound shows $\Universal(\cA, (P, R)) \in \Omega(\log |\bH_{P, R}|)$.
    Finally, we consider the witnesses.
    A list of integers $W \in \mathbb{Z}^{n \times 3}$ is a witness for at most $\Vmax(P, R)$ different inputs in $\bI_{P, R}$.
    Thus, the algorithm $\cA$ needs to output at least $\lceil \frac{|\bI_{P, R}|}{\Vmax(P, R)}\rceil$
    different lists $W$ across all inputs in $\bI_{P, R}$.
    The information theoretic lower bound shows $\Universal(\cA, (P, R)) \in \Omega(\log \frac{|\bI_{P, R}|}{\Vmax(P, R)})$.
\end{proof}

\subsection{Quadrangle Tree and Ordered Downdrafts}

For the Pareto front, there was a natural partial order on $P$ given by domination.
We then used this partial order to define downdrafts.
Moreover, \cref{lemm:pareto_linear_order} shows that this partial order goes hand-in-hand with regions that are sorted by $x$- or $y$-coordinate.
For the convex hull problem, we can use the quadrangle tree~\cite{van_der_hoog_combinatorial_2025} to get a partial order on $P$.
However, we will need to permute this partial order to make it compatible with how regions are sorted.
Downdrafts are defined a lot less geometric.

\begin{definition}[\cite{van_der_hoog_combinatorial_2025}]
    A rooted convex polygon $(C, p, q)$ is a convex polygon $C$ together with an edge $pq$ of $C$. Given distinct vertices $r, s$ of $C$, the line $rs$ splits $C$ into two pieces (one piece is empty if $rs$ is an edge of $C$). Let $C^{rs}$ denote the piece that does not contain $pq$.
\end{definition}

\begin{definition}[\cite{van_der_hoog_combinatorial_2025}]
    A quadrangle tree $\cQ$ of a rooted convex polygon $(C, p, q)$ is a binary tree in which every node stores a quadrangle spanned by (up to) four vertices of $C$, such that:
    \begin{itemize}[noitemsep]
        \item the quadrangle of the root node is spanned by $p, q, s, r$ where $rs$ is an edge of $C$.
        
        We allow for $p=r$ and/or $q=s$.
        \item If $q \ne r$, then the root has  a child node whose subtree is a quadrangle tree of $(C^{p r}, p, r)$.
        \item If $s \ne p$, then the root has  a child node whose subtree is a quadrangle tree of $(C^{q s}, q, s)$.
    \end{itemize}
    Given a point set $P$ and a quadrangle tree $\cQ$, the \emph{population} of the root node are all points in $P$ that lie on or inside the quadrangle $pqrs$, except for those that lie on the line $pq$. 

    Given a region set $R$, a \emph{truncated quadrangle tree} moreover makes a node a leaf if all remaining points $P'$ are contained in a single region $r \in R$. The population of such a leaf consists of all points in $P'$.
\end{definition}

\begin{definition}[\cite{van_der_hoog_combinatorial_2025}]
    We define for any (truncated) quadrangle tree $\cQ$ of $\CH(P)$ a partial order $\prec_\cQ$ on $P$. 
    For $p \in P$, let $r(p)$ be the node in $\cQ$ within whose population $p$ lies.
    Let $H_{r(p)}$ be the halfplane defined by the rooted edge of $r(p)$. 
    For $p, q \in P$, we say that $p \prec_\cQ q$ if
    \begin{itemize}[noitemsep]
        \item $r(p)$ is a strict ancestor of $r(q)$, or
        \item $r(p) = r(q)$, and $q$ lies deeper inside the halfspace $H_{r(p)}$ than $p$.
    \end{itemize}
\end{definition}

\begin{lemma} \label{lemm:quadrangle_order}
    Let $P$ be a point set and $R$ a region set. Let $\cQ$ be a (truncated) quadrangle tree of $\CH(P)$.
    Let $\kappa \in \CF(P, R)$ be a compass function from Definition~\ref{def:compass}.
    There is a permutation $\rho \in S_P$ such that, for any points $p, q \in P$
    \begin{enumerate}[(1), noitemsep]
        \item if $u(p) \ne u(q)$, then $\rho(p) \prec_{\cQ} \rho(q)$ if and only if $p \prec_{\cQ} q$.
        \item if $u(p) = u(q)$ and $p, q$ lie in the same region $r \in R$, then $\rho(p) \prec_{\cQ} \rho(q)$ is the sorted order by $\kappa(r)$.
    \end{enumerate}
    We denote by $\prec_{\cQ, \kappa}$ the partial order defined via: $p \prec_{\cQ, \kappa} q$ if and only if $\rho^{-1}(p) \prec_{\cQ} \rho^{-1}(q)$.
    And by $<_{\cQ, \kappa}$ an arbitrary linear extension of $\prec_{\cQ, \kappa}$.
\end{lemma}
\begin{proof}
    For a node $u \in \cQ$ and a region $r \in R$, let $P_{u, r} = \set{p \in P \cap r \mid u(p) = u}$.
    Let $p_1 \prec_{\cQ} \dots \prec_{\cQ} p_k$ be the points in $P_{u, r}$ ordered by $\prec_{\cQ}$,
    and let $q_1, \dots, q_k$  be the same set of points ordered by $\kappa(r)$.
    We put $\rho(q_i) = p_i$ for $i \in [k]$.
    We do this construction for every pair $(u, r)$.
    Since the sets $P_{u, r}$ form a partition of $P$, this defines a permutation $\rho \in S_P$.
    
    Let $p, q \in P$ be points. If $u(p) \ne u(q)$, then $p \prec_{\cQ} q$ only depends on $u(p)$ and $u(q)$,
    and not on the points themselves. Since $u(\rho(p)) = u(p)$ and $u(\rho(q)) = u(q)$, 
    we have $\rho(p) \prec_{\cQ} \rho(q)$ if and only if $p \prec_{\cQ} q$. This shows (1).
    Finally, we consider points inside the same set $P_{u, r}$. Define $p_1, \dots, p_k$ and $q_1, \dots, q_k$ as above.
    We have $\rho(q_i) \prec_{\cQ} \rho(p_j)$ if and only if $i < j$. But also $i< j$ if and only if $q_i$ comes before $q_j$ in the ordering by $\kappa(r)$. This shows (2).
\end{proof}

\subparagraph{Normalized Downdrafts} Let $P$ be a point set, $R$ a region set, and $\kappa \in \CF(P, R)$ a compass function. Let $\cQ$ be a (truncated) quadrangle tree.
A \emph{downdraft} is a map $\varphi : P - \ch(P) \to P$ such that for each $p \in P - \ch(P)$, we have $p \prec_{\cQ, \kappa} \varphi(p)$. A \emph{normalized downdraft} is a downdraft $\psi$ such that for all regions $r \in R$ and points $p, q \in P \cap r$ with $p <_{\cQ, \kappa} q$, we have $\psi(p) \leq_{\cQ, \kappa} \psi(q)$.

\subparagraph{Ordered Normalized Downdrafts.} 
Given any normalized downdraft $\psi$ and point $q \in P$, the \emph{fiber} $\psi^{-1}(\{q\})$ is the set of all points $p \in P$ that get mapped to $q$.
We define the \emph{ordered normalized downdraft} $\ol{\psi} = (\psi, \prec_{\ol{\psi}})$  as a normalized downdraft $\psi$ plus 
a total order $\prec_{\ol{\psi}}$ on each fiber $\psi^{-1}(\{p\})$,
such that for all regions $r \in R$ and points $p, q \in P \cap r$ with $\psi(p) = \psi(q)$,
we have $p <_{\prec_{\cQ, \kappa}} q \Leftrightarrow p \prec_{\ol{\psi}} q$.
Similarly to how we count witness sets, we denote by $\OND(P, R, \cQ, \kappa)$ the set of all ordered normalized downdrafts and $\ONDmax(P, R, \cQ) = \max\limits_{\kappa \in \CF(P, R)} |\OND(P, R, \cQ, \kappa)|$.

\subsection{The Eppstein-Goodrich-Illickan-To Algorithm for Convex Hull}

Kirkpatrick, McQueen and Seidel design a convex hull algorithm~\cite{kirkpatrick1986ultimate}.
Eppstein, Goodrich, Illickan and To~\cite{eppstein_entropy-bounded_2025}
design a truncated version of the this algorithm that terminates early if all remaining points are sorted, e.g., by x-coordinate, see \cref{algo:convex_hull}.
We upper bound the running time of their algorithm by the quantities: $\sum\limits_{p\in P} \depth_{Q(P,R)}(u(p))$, $\ONDmax(P, R, Q(P, R)) $ and $|\mathbb{I}_{P, R}|$ exactly as we did in Section~\ref{sub:counting}.

\begin{algorithm}[h] 
\caption{Eppstein-Goodrich-Illickan-To-Hull$(I', \textnormal{ edge } (p_\ell, p_r) \textnormal{ of } \CH(I') )$ \cite{kirkpatrick1986ultimate}.} \label{algo:convex_hull}
\begin{algorithmic}[1]
    \REQUIRE $\set{p \in I'} = \set{p \in P \ | \ p \text{ lies on or above the line } p_\ell p_r}$ and $p_\ell, p_r \in \ch(P)$.
    \IF{$I'$ is sorted}
        \STATE Compute $\ch(I')$ and a witness for each point not on $\ch(I')$ in linear time
        \STATE Update the hull list and witness list
        \STATE \textbf{return}
    \ENDIF
    \STATE $m \gets \arg \operatorname{median}\set{x(p) \ | \ p \in S}$
    \STATE $(p_i, p_j) \gets $ edge of $\CH(S)$ such that $m \in Q = (p_\ell, p_r, p_j, p_i)$  \hspace{1.2cm }\emph{bridge-finding}~\cite{kirkpatrick1986ultimate}.
    \STATE $S_1 \gets \set{p \in S \ | \ p \text{ lies on or above the line } p_\ell p_i}$
    \STATE $S_2 \gets \set{p \in S \ | \ p \text{ lies on or above the line } p_j p_r}$
    \STATE $S^* \gets S - S_1 - S_2$
    \STATE $\operatorname{Eppstein-Goodrich-Illickan-To-Hull}(S_1, (p_\ell, p_i))$
    \STATE $\operatorname{Eppstein-Goodrich-Illickan-To-Hull}(S_2, (p_j, p_r))$
     \STATE GiveWitness($S^*$, $p_\ell$, $p_r$, $p_j$, $p_i$) \hspace{2cm} \emph{Note that this last step is not explicit in~\cite{kirkpatrick1986ultimate}.}
\end{algorithmic}
\end{algorithm}

The execution of \cref{algo:convex_hull} depends on the order in which P appears in I.
Let $R$ be a region set. We modify Algorithm 2 by replacing the condition ``$I'$ is sorted'' with ``$P'$ lies entirely in some region $r \in R$''. The resulting algorithm depends only on $P$ and $R$.
Its recursion tree is a truncated quadrangle tree $Q(P, R)$, with each node corresponding to a recursive call.
The quadrangle of this node is $p_{\ell} p_i p_j p_r$.

\begin{lemma} \label{lemm:ch_running_time}
Fix a universe $(P,R)$ and let $I \in \bI_{P,R}$.
Let $Q(P,R)$ be the truncated quadrangle tree induced by the modified \cref{algo:convex_hull}.
The running time of the original \cref{algo:convex_hull} on $I$ is
$O\!\left(n + \sum\limits_{p\in P} \depth_{Q(P,R)}(u(p))\right)$,
where $\depth_{Q(P,R)}(u(p))$ is the depth of $u(p)$ in $Q(P,R)$.
\end{lemma}
\begin{proof}
The proof is analogous to that of \cref{lemm:pf_running_time}.
\end{proof}

A region $r \in R$ might intersect the population of many nodes in the quadrant tree.
We again define a region set $R'$ such that every region in $r'$ intersects the population of at most one node in $Q(P, R)$.
Observe that this refinement step does not change the truncated quadrangle tree. Therefore, $Q(P, R) = Q(P, R')$.

\begin{lemma} \label{lemm:quadrangle_tree_ond}
    Fix the universe $(P, R)$ and let $R'$ be a region set of the same area where every region $r' \in R'$ intersects the population of at most one node in $Q(P, R)$.
    Consider the truncated quadrangle tree $Q(P, R')$ induced by running the modified Algorithm~\ref{algo:convex_hull} on $(P, R')$. 
    Then
    \[
        \ONDmax(P, R', Q(P, R')) \le   \ONDmax(P, R, Q(P, R))  \leq \frac{n^n}{2^{\sum_{p \in P} \left( \depth_{Q(P, R')} \cdot \; u(p) \right)}} \cdot \prod_{r \in R'} \frac{1}{|r \cap P|!}
    \]
\end{lemma}
\begin{proof}
    The proof is identical to that of \cref{lemm:quadrant_tree_ond},
    except that the linear order $<_L$ gets replaced by the linear order $<_{Q(P, R'), \kappa}$ defined in \cref{lemm:quadrangle_order}.
\end{proof}

\begin{theorem} \label{theo:hull_upper_bound}
    The running time of \cref{algo:convex_hull} is $
        O\Big(n + \log \frac{|\bI_{P, R}|}{\ONDmax(P, R, Q(P, R))} \Big).$
\end{theorem}
\begin{proof}
    By \cref{lemm:ch_running_time}, the running time is bounded by  $
        O\Big(n + \sum\limits_{p \in P} \depth_{Q(P, R)} \left( u(p) \right) \Big)$.
        
    As noted earlier, $Q(P, R') = Q(P, R)$. We apply \cref{lemm:quadrangle_tree_ond} and obtain:
    \[
        \sum_{p \in P} \depth_{Q(P, R)} u(p) \le \log \Big(\frac{n^n}{\ONDmax(P, R', Q(P, R'))} \cdot \prod_{r \in R'} \frac{1}{|r \cap P|!} \Big).
    \]
    Since $R'$ is a refinement of $R$, we have $\ONDmax(P, R', Q(P, R')) \ge \ONDmax(P, R, Q(P, R'))$.
    We apply an algebraic argument identical to the last step of \cref{lemm:quicksort_upper} to derive the following upper bound:
    
    \[
        \prod_{r \in R'} \frac{1}{|r \cap P|!} \le 2^{O(n)} \cdot \prod_{r \in R} \frac{1}{|r \cap P|!}. \hfill \text{ By Stirling's approximation, $n^n \le n! \cdot 2^{O(n)}$.
    Therefore:}
    \]

    \[
        \sum_{p \in P} \depth_{Q(P, R)} u(p) \le \log \Big(\frac{1}{\ONDmax(P, R, Q(P, R))} \cdot \frac{n!}{\prod_{r \in R} |r \cap P|!}\Big) + O(n)
    \]
    Finally, $\bI_{P, R}$ contains all inputs for which the points in every region are sorted by increasing x-coordinate, thus
    \[
        \frac{n!}{\prod\limits_{r \in R} |r \cap P|!} \le |\bI_{P, R}|. \qedhere
    \]
\end{proof}

\subsection{The Counting Argument for Convex Hull}

What remains is to show that the lower bound $\Omega\Big(n + \log |\bH_{P, R}| + \log \frac{|\bI_{P, R}|}{\Vmax(P, R)}\Big)$ matches the upper bound $
        O\Big(n + \log \frac{|\bI_{P, R}|}{\ONDmax(P, R, Q(P, R))} \Big)$.
        Just as in Section~\ref{sect:pareto}, we show this relation via an injective map $\Phi$. 

\begin{definition}
    Let $W = (a_i, b_i, c_i)_{i=1}^{n}$.
    A \emph{corner assignment}~\cite{van_der_hoog_combinatorial_2025} picks a corner for each of the $n$ triangles in $W$. 
    Formally, it is a map $C : [n] \to [n]$ such that
    $C(i) \in \{a_i, b_i, c_i\}$ for all $i \in [n]$.
    Let $\CA(W)$ denote the set of all corner assignments of $W$. Note that $|\CA(W)| = 3^n$.
\end{definition}

\begin{theorem}
    \label{thm:hull_map}
    Fix the universe $(P, R)$ and list $W  = (a_i, b_i, c_i)_{i=1}^{n} \in \mathbb{Z}^{n \times 3}$. 
    Let $\cQ$ be an arbitrary (truncated) quadrangle tree.
    Recall that  $V(P, R, W)$ denotes the inputs $I \in \bI_{P, R}$ for which $W$  is a witness list for $I$.
    There is a natural map
    \[
        \Phi : V(P, R, W) \hookrightarrow \bigcup_{\kappa \in \CF(P, R)} \Big(\{\kappa\} \times \OND(P, R, \cQ, \kappa) \times \bH_{P, R} \times \CA(W)\Big), %\hfill 
    \]

and this map is injective.
\end{theorem}

\begin{figure}[h]
\centering
    \includegraphics[page=2]{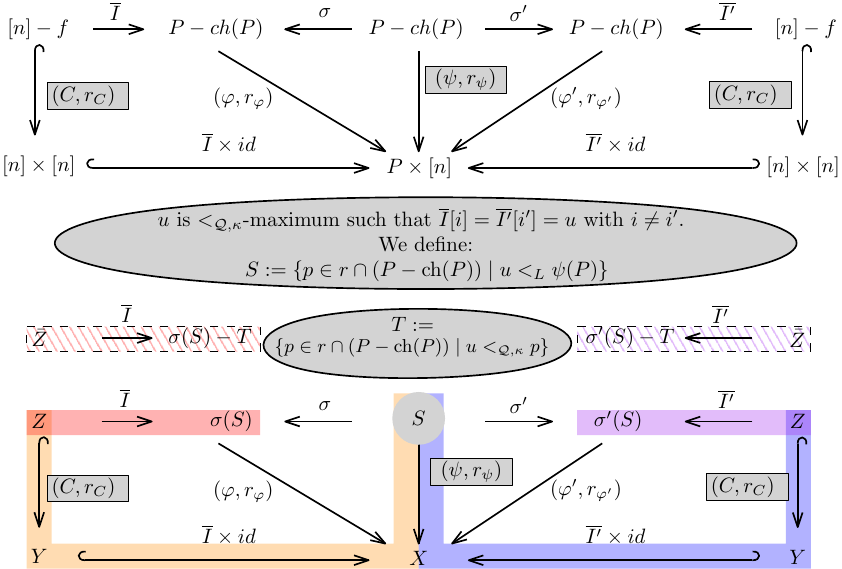}
    \caption{Top: the commuting diagram. Grey elements are `shared' between the distinct inputs $I$ and $I'$.
    Bottom: the diagram chasing, starting from $S$.} \label{fig:hull_commuting_diagram}
\end{figure}

\begin{proof}
Let $I \in V(P, R, W)$. We view $I$ as a map $[n] \mapsto P$ and we construct $\Phi$:

\paragraph{Defining the map $\Phi$.}
There is a unique hull list $h \in \bH_{P, R}$ for $I$.
We define a compass function $\kappa$ as follows: for each $r \in R$,
$\kappa(r)$ describes the order in $I \cap r$.
E.g., whether the points in $P \cap r$ appear in $I$ sorted by $x$-coordinate.
If the input is sorted increasingly by both $x$ and $y$, then $\kappa(r)$ chooses sorting by $x$.

We define a corner assignment $C$ as follows.
For every $i \in [n]$ with $I[i] \notin \ch(P)$, the point $I[i]$ lies strictly inside the triangle $\Delta(I[a_i], I[b_i], I[c_i])$.
Therefore, at least one of $I[i] \prec_{\cQ} I[a_i]$, $I[i] \prec_{\cQ} I[b_i]$, or $I[i] \prec_{\cQ} I[c_i]$ holds.
We let $C(i)$ be the first index in the order $a_i,b_i,c_i$ for which this relation holds.
For every $i \in [n]$ with $I[i] \in \ch(P)$, we put $C(i)=-1$.

Let $\rho$ be the permutation from \cref{lemm:quadrangle_order}.
We define $\ol I=\rho \, \circ \, I$, that is,
$\ol I=(\rho(I[1]),\ldots,\rho(I[n]))$.
We define a downdraft $\varphi$ as follows:
for every $i \in [n]$ with $C(i)\neq-1$, put
$\varphi(\ol I[i])=\ol I[C(i)]$.
By the definition of $C$, we have $I[i]\prec_{\cQ}I[C(i)]$.
Hence, by the definition of $\prec_{\cQ,\kappa}$ in \cref{lemm:quadrangle_order},
$\ol I[i]=\rho(I[i])\prec_{\cQ,\kappa}\rho(I[C(i)])=\ol I[C(i)]$.
Thus, $\varphi$ is a downdraft.

    Next, we define a permutation $\sigma$ that permutes the points inside each region,
    and use $\sigma$ to turn $\varphi$ into a normalized downdraft $\psi$.
    Recall the linear order $<_{\cQ, \kappa}$ defined in \cref{lemm:quadrangle_order}.
    For any region $r$, let $p_1 <_{\cQ, \kappa} \dots <_{\cQ, \kappa} p_k$ be the points in $r \cap (P - \ch(P))$ sorted by $<_{\cQ, \kappa}$,
    and let $q_1, \dots, q_k$ be the same set of points, but sorted by $\varphi(q_1) \le_{\cQ, \kappa} \dots \le_{\cQ, \kappa} \varphi(q_k)$.
    If there are ties, say $\varphi(q_i) = \varphi(q_{i+1})$ for some index $i$, then we require that $q_i <_{\cQ, \kappa} q_{i+1}$.
    We put $\sigma(p_i) = q_i$ and $\psi = \varphi \circ \sigma$.

    Finally, we define an ordered normalized downdraft $\ol{\psi} = (\psi, <_{\ol{\psi}})$:
    Let $i, j \in [n]$ with $\varphi(\ol{I}[i])) = \varphi(\ol{I}[j]))$.
    Then $\psi(\sigma^{-1}(\ol{I}[i])) = \varphi(\rho(\ol{I}[i])) = \varphi(\rho(\ol{I}[j])) = \psi(\sigma^{-1}(\rho(\ol{I}[j])))$,
    so $\sigma^{-1}(\rho(\ol{I}[i]))$ and $\sigma^{-1}(\rho(\ol{I}[j]))$ lie in the same fiber of $\psi$.
    We put $\sigma^{-1}(\ol{I}[i]) \prec_{\ol{\psi}} \sigma^{-1}(\ol{I}[j])$ if and only if $i < j$.
    Put $\Phi(I) = (\kappa, \ol{\psi}, f, C)$.
    By doing this for every $I\in V(P,R,W)$, we get for each input $I$
a compass function, an ordered normalized downdraft, the unique hull list
$h$, and a corner assignment, which defines the map $\Phi$.

    \paragraph{Showing a contradiction.}
    For any $I \in V(P, R, W)$, our definition of $\Phi$ constructs some pair $(\sigma, \varphi)$ which leads to the output of $\Phi(I)$. Similarly, any $I' \in V(P, R, W)$ creates some pair $(\sigma', \varphi')$ which leads to $\Phi(I')$. Suppose for the sake of contradiction that $\Phi$ is not injective.
    Then for distinct $I, I' \in V(P, R, W)$, $\Phi(I) = \Phi(I') = (\kappa, \ol{\psi}, h, C)$.
    We immediately note that these two inputs $I$ and $I'$ `share' $P$, $R$, $\kappa$, $\ol{\psi}$, $h$, $C$, the permutation $\rho$ defined in \cref{lemm:quadrangle_order} and the linear order  $<_{\cQ, \kappa}$ defined in \cref{lemm:quadrangle_order} as all these items are defined by $P, R$ and $\kappa$ (where $P$ and $R$ are shared by definition and $\kappa$ by assumption of the map outputs being equal). 
     A priori, we do not know whether $\sigma = \sigma'$ or $\varphi = \varphi'$.
However, we can conclude the following fact from the construction of $\Phi$.
For any index $i\in[n]-h$, we have
\[
\psi(\sigma^{-1}(\ol I[i]))=\varphi(\ol I[i])=\ol I[C(i)]
\quad \quad \text{and} \quad \quad
\psi(\sigma'^{-1}(\ol I'[i]))=\varphi'(\ol I'[i])=\ol I'[C(i)]. \]

 Let $<_{\cQ,\kappa}$ be the linear order from \cref{lemm:quadrangle_order}.
Since $I\neq I'$, there are distinct integers $i,i'\in[n]$ with
$I[i]=u=I'[i']$.
Pick $(i,i')$ so that $u$ is maximal with respect to $<_{\cQ,\kappa}$.
We first consider the case where $u\in\ch(P)$, say, $u$ is the $k$'th point on $\ch(P)$.
Since $\Phi(I)=\Phi(I')=(\kappa,\ol{\psi},h,C)$, both inputs share the same hull list $h$.
But then $I[h[k]]=u=I'[h[k]]$, which implies $i'=h[k]=i$, a contradiction.
Therefore, $u\notin\ch(P)$.

\paragraph{Creating a commuting diagram.}
The remainder of our proof relies on a commuting diagram. 
     We define the shared functions $r_C$ and $r_\psi$, and functions $r_\varphi$ and $r_{\varphi'}$ for $\ol I$ and $\ol I'$, respectively (see Figure~\ref{fig:hull_commuting_diagram}). 
     Both inputs have the same corner assignment $C$. 
Let $r_C$ assign every $i\in[n]-h$ its rank in the fiber $C^{-1}({C(i)})$, when this fiber is ordered increasingly by index.
Similarly, let $r_\psi$ assign every $u\in P-\ch(P)$ its rank in the fiber $\psi^{-1}({\psi(u)})$, when this fiber is ordered by $\prec_{\ol\psi}$.
We define $r_\varphi$ by ordering each fiber of $\varphi$ according to the positions of its points in $\ol I$, and define $r_{\varphi'}$ analogously using $\ol I'$.

By the $\prec_{\ol\psi}$-part of the definition of $\Phi$, for every $i\in[n]-h$, we have
$r_\psi(\sigma^{-1}(\ol I[i]))=r_\varphi(\ol I[i])=r_C(i)$.
Similarly,
$r_\psi(\sigma'^{-1}(\ol I'[i]))=r_{\varphi'}(\ol I'[i])=r_C(i)$.
Thus, the diagram in Figure~\ref{fig:hull_commuting_diagram} commutes.

    \paragraph{Chasing our commuting diagram.}
    Let $u \not \in ch(P)$ and let $r \in R$ be the region that contains $u$. 
    We define the set $S := \set{p \in r \cap (P - \ch(P)) \mid u <_{\cQ, \kappa} \psi(p)}$.
    We now chase the diagram in \cref{fig:hull_commuting_diagram}.
We now chase the diagram, starting with the orange path in \cref{fig:hull_commuting_diagram}.
Starting from $S$, we apply $(\psi,r_\psi)$ to obtain a set $X\subseteq P\times[n]$,
then apply $(\ol I\times\id)^{-1}$ to obtain a set $Y\subseteq[n]\times[n]$,
and finally apply $(C,r_C)^{-1}$ to obtain a set $Z\subseteq[n]-h$.
Since the diagram commutes and the maps $(\ol I\times\id)$, $(C,r_C)$, and $(\psi,r_\psi)$ are injective, we have
$\sigma^{-1}(\ol I[Z])=S$.

We can alternatively apply the blue path in \cref{fig:hull_commuting_diagram}. That is, starting from $S$ and applying the shared pair $(\psi,r_\psi)$ to obtain $X$.
We then apply $(\ol I'\times\id)^{-1}$ and $(C,r_C)^{-1}$.
For every $v\in\psi(S)$, we have $u<_{\cQ,\kappa}v$ by the definition of $S$.
Recall that $u$ was chosen as the $<_{\cQ,\kappa}$-maximal point such that there exist distinct $i$ and $i'$ where $I[i] = u = I'[i]$. 
Therefore, every $v$ with $u<_{\cQ,\kappa}v$ appears at the same index in $I$ and $I'$. 
Since both inputs `share' the same permutation $\rho$, this also means that all inputs $v$ with $u<_{\cQ,\kappa}v$ appears at the same index in $\ol{I}$ and $\ol{I'}$. 
%HOWEVER, in $\ol{I}$ and $\ol{I'}$ the same two indices, can lead to a different point!
This holds in particular for every $v\in\psi(S)$.
$(\ol I\times\id)^{-1}(X)=(\ol I'\times\id)^{-1}(X)$,
so both the orange and the blue routes through the diagram yield the same sets $Y$ and $Z$.
    In particular, we now have a set $Z$ with $\ol{I}[Z] = \sigma(S)$ and $\ol{I}'[Z] = \sigma'(S)$.

We now define the set
$T := \{p \in r \cap (P-\ch(P)) \mid u <_{\cQ,\kappa} p \}$.
Note that the definition of $T$ is independent of $I$ and $I'$.
We have
$\sigma(S)= \{p\in r\cap(P-\ch(P))\mid u<_{\cQ,\kappa}\varphi(p) \}$
by the definition of $S$.
For every $p\in T\cup{u}$, we have
$p\prec_{\cQ,\kappa}\varphi(p)$, since $\varphi$ is a downdraft, and hence
$p<_{\cQ,\kappa} \varphi(p)$.
Since $u\le_{\cQ,\kappa }p$, it follows that
$u<_{\cQ,\kappa}\varphi(p)$, so $p\in\sigma(S)$.
Therefore,
$T\cup{u}\subseteq\sigma(S)$ and, via the same argument, $T\cup{u}\subseteq\sigma'(S)$.

    By maximality of $u$, we have $\ol{I}^{-1}(T) = \ol{I}'^{-1}(T)$.
    Put $\ol{Z} = Z - \ol{I}^{-1}(T)$. Recall that $W$ is the witness list. Then $\ol{I}[\ol{Z} ] = \sigma(S) - T$ and $\ol{I}'[\ol{Z} ] = \sigma'(S) - T$.
    Observe that $u$ is the $<_{\cQ, \kappa}$-maximal point in $\sigma(S) - T$ and in $\sigma'(S) - T$,
    and that $\sigma(S) - T \subseteq r$ and $\sigma'(S) - T \subseteq r$.
    By part (2) of \cref{lemm:quadrangle_order}, this means that $i = \max(\ol{Z}) = i'$.
    This contradiction shows that $\Phi$ is injective.
\end{proof}

\begin{corollary} \label{coro:hull_counting}
    Let $P$ be a point set and $R$ a region set.
    Let $\cQ$ be any (truncated) quadrangle tree.
    Then
    \[
        \Vmax(P, R) \le |\bH_{P, R}| \cdot \ONDmax(P, R, \cQ) \cdot 4^{n} \cdot 3^{n}.
    \]
\end{corollary}
\begin{proof}
    Regions $r \in R$ that contain no points of $P$ have no effect on $\Vmax(P, R)$ or $\OND(P, R)$.
    Thus, we may assume that every region in $R$ contains a least one point in $P$,
    so $|R| \le n$.
    By Theorem~\ref{thm:pareto_map}, for every witness $W$
    \begin{align*}
        V(P, R, W) &\le \sum_{\kappa \in \CF(P, R, W)} |\OND(P, R, \kappa, \cQ)| \cdot |\bH_{P, R}| \cdot |\CA(W)|\\
        &\le |\CF(P, R, W)| \cdot \ONDmax(P, R, \cQ) \cdot |\bH_{P, R}| \cdot |\CA(W)|.
    \end{align*}
    Since $|\CF(P, R, W)| \le 4^{|R|} \le 4^n$ and $|\CA(W)| \le 3^n$,
    the corollary follows by taking the maximum over $W$.
\end{proof}

\convexTHM*

\begin{proof}[Proof of Theorem~\ref{thm:convex} (i)]
    By \cref{theo:hull_upper_bound}, the universal running time of \cref{algo:pareto_front} is $
        O\Big(n + \log \frac{|\bI_{P, R}|}{\ONDmax(P, R, Q(P, R))}\Big)$.
    By \cref{coro:hull_counting}, 
    \[
        \ONDmax(P, R, Q(P, R)) \ge \frac{\Vmax(P, R)}{|\bH_{P, R}| \cdot  4^{n} \cdot 3^{n}},
    \]
    so the running time of \cref{algo:convex_hull} is
    \[
        O\Big(n + \log \frac{|\bI_{P, R}|}{\Vmax(P, R)} + \log |\bH_{P, R}|\Big), \quad \text{ which shows universal optimality.}\qedhere
    \]
\end{proof}

\subsection{A Generalization of ``Sorted by x- or y-coordinate''}
\label{sec:generalization}

In \cref{sect:pareto}, it was it was important that the compass function sorts points by x-coordinate or y-coordinate. Indeed, \cref{lemm:pareto_linear_order} constructs a linear order that both compatible with the domition partial order, and with the compass function.
In contrast to this, the proofs in \cref{sect:hull} work for a wide range of compass functions. Indeed, the permutation $\rho$ in \cref{lemm:quadrangle_order} can deal with an arbitrary compass function.
Thus, we may consider a bigger family of compass functions that just ``sorted by x- or y-coordinate''.

For \cref{sect:hull}, a family of compass functions really only needs to satisfy three properties:
\begin{enumerate}[(1)]
\item We need to decide in linear time whether there exists a compass function according to which $I'$ is sorted.
\item If $I'$ is sorted via some compass function, then we need an algorithm that computes the convex hull of $I'$ in linear time.
\item The number of distinct compass functions should be $2^{O(n)}$.
\end{enumerate}

Indeed, property (1) is used on line 1 of \cref{algo:convex_hull}, and property (2) is used in line 2 of \cref{algo:convex_hull}. Property (3) is needed for the analysis, since the number of compass functions shows up in \cref{coro:hull_counting}. (Also, if all points in $R$ lie on the convex hull, then property (2) implies property (3) via an information theoretic lower bound.)

\subparagraph{Generalized Compass Functions} A \emph{generalized compass function} $\kappa$ assigns every region $r \in R$
an order $p_1, p_2, \dots, p_k$ of the points in $r \cap P$, such that $p_1 p_2 \dots p_k$ is a simple polygonal curve. In other words, $p_1 p_2 \dots p_k$ is a (directed) spanning path of the point set $r \cap P$.
This generalizes the compass functions from \cref{def:compass}: If $p_1.x < p_2.x < \dots < p_k.x$,
then $p_1 p_2 \dots p_k$ is a simple polygonal curve. This also holds for sorting by y-coordinate and/or in decreasing order.

Let us check that this definition satisfies the conditions (1)-(3):
\begin{enumerate}[(1)]
\item Deciding whether the points in $I'$ form a simple polygonal curve can be done in $O(|I'|)$ time via Chazelle's polygon triangulation algorithm (second application in Section 5 of~\cite{chazelle_triangulating_1991}).
\item The convex hull of a simple polygonal curve can be computed in linear time via a variant of Graham scan~\cite{lee1983finding}.
\item Any planar point set of size $k$ has $O(86.81^k)$ spanning paths \cite{sharir_number_2006}\footnote{The abstract of~\cite{sharir_number_2006} only mentions spanning cycles, but at the end of Section 5.1 in \cite{sharir_number_2006}, the authors note that their bound also applies to spanning paths.}. Let $c>0$ be a constant so that this bound is $\le c \cdot 86.81^k$ for every integer $k \ge 0$. Then, for a single region $r \in R$, there are $\le c \cdot 86.81^{|r \cap P|}$ generalized compass functions.
By taking the product over all regions that contain at least one point of $P$, the total number of generalized compass functions is $\le c^n \cdot 86.81^n \in 2^{O(n)}$.
\end{enumerate}

\section{A Lower Bound Based on Entropy}
\label{sec:entropy}

As a byproduct of our results, we will derive a lower bound based on the entropy of \cite{eppstein_entropy-bounded_2025}.
We spell this out in detail for the Pareto front. The proof for quicksort and convex hull are similar.
Recall that  Eppstein, Goodrich, Illickan, and To~\cite{eppstein_entropy-bounded_2025} show that,
on any input $I$, \cref{algo:convex_hull} has a running time of $O(n (1+H(I))$,
where $H$ denotes the range partition entropy. This raises the question of proving a matching lower bound.

Showing, for every algorithm $\cA$, a lower bound of $\Runtime(\cA, I) \in \Omega(n (1+H(I))$ for every individual input $I$ is impossible.
Indeed, such a bound would imply that \cref{algo:convex_hull} is \emph{instance optimal}.
But Afshani, Barbay, and Chan~\cite{afshani2009instance} show that instance optimal convex hull algorithms do not exist. 
We thus pursue and achieve the next best option: Showing a lower bound of $\Universal(\cA, (P, R)) \in \Omega(n (1+H(I)))$ for every universe $(P, R)$ that contains $I$.
This is precisely the statement $\UniversalLB(\dots) \in \max_{I \in \dots} \Omega(n(1+H(I)))$
in \cref{thm:quicksort,thm:pareto,thm:convex} part (ii).
The idea is to show a $\Omega(n(1+H(I))$ lower bound for the algorithm from \cite{eppstein_entropy-bounded_2025}, and then use universal optimality to extend the lower bound to any algorithm.

\begin{lemma} \label{lemm:quick_slow}
    Let $\Aquick$ denote the sorting algorithm from \cite{eppstein_entropy-bounded_2025}.
    Then $\Runtime(\Aquick, I) \in \Omega(n(1+H(I))$ for any any input $I$ of length $n$.
\end{lemma}

\begin{proof}[Proof of \cref{thm:quicksort} part (ii) using \cref{lemm:quick_slow}]
    Let $I \in \bI_{s_1, \dots, s_k}$ be arbitrary. By part (ii) of \cref{thm:quicksort}
    \[
        \UniversalLB(s_1, \dots, s_k) \ge \frac{1}{c} \Universal(\Aquick, (s_1, \dots, s_k)).
    \]
    Since $I \in \bI_{s_1, \dots, s_k}$ and by \cref{lemm:algo_slow},
    \[
        \Universal(\Aquick, (s_1, \dots, s_k)) \ge \Runtime(\Aquick, I) \in \Omega(n(1+H(I))).
    \]
    Therefore, $\UniversalLB(s_1, \dots, s_k) \in \Omega(n(1+H(I))$ for all $I \in \bI_{s_1, \dots, s_k}$.
    Part (ii) of the Theorem follows by taking the maximum over all such $I$.
\end{proof}

The proof of \cref{lemm:quick_slow} involves constructing a respectful partition from the recursion tree of $\Aquick$,
plus some tedious but routine algebraic manipulations.
We illustrate this for the Pareto front, see \cref{lemm:algo_slow}.
The proof for quicksort is similar.

\subsection{Pareto front}

We now show part (ii) of:

\paretoTHM*

We first show this lower bound for \cref{algo:pareto_front}. Then we use the universal optimality of \cref{algo:pareto_front} to extend the lower bound to any algorithm.

\begin{lemma} \label{lemm:algo_slow}
    Let $\Apf$ be \cref{algo:pareto_front}.
    Then $\Runtime(\Apf, I) \in \Omega(n(1+H(I)))$ for any input $I$.
\end{lemma}

\begin{proof}[Proof of \cref{thm:pareto} part (ii) using \cref{lemm:algo_slow}]
    Let $I \in \bI_{P, R}$ be arbitrary.
    By part (ii) of \cref{thm:quicksort}
    \[
        \UniversalLB(P, R) \ge \frac{1}{c} \Universal(\Apf, (P, R)).
    \]
    Since $I \in \bI_{P, R}$ and by \cref{lemm:algo_slow},
    \[
        \Universal(\Apf, (P, R)) \ge \Runtime(\Apf, I) \in \Omega(n(1+H(I))).
    \]
    Part (ii) of the Theorem follows by taking the maximum over all $I$.
\end{proof}

\begin{proof}[Proof of \cref{lemm:algo_slow}]
    Let $I$ be any input. We will construct a suitable respectful partition $\Pi = ((R^1, I^1), \dots)$. (We use raised indices to avoid confusion with the sets $I_1, I_2$ in \cref{algo:pareto_front}.)
    Consider the execution of \cref{algo:pareto_front} on $I$.
    Whenever \cref{algo:pareto_front} returns because $I'$ is sorted,
    put $I^i = I'$ and let $R^i$ be the bounding box of $I'$.
    Whenever \cref{algo:pareto_front} computes $I_1, I_2$,
    let $I^i = I' - (I_1 \cup I_2)$ and let $R^i$ be the bounding box of $I^i$ (which is dominated by the point $q$ in \cref{algo:pareto_front}).
    Due to how \cref{algo:pareto_front} constructs the sets $I_1, I_2$,
    all the $R^i$ are disjoint.
    Thus, $\Pi = ((R^1, I^1), \dots)$ is indeed a respectful partition for $I$.
    Moreover, each $(R^i, I^i)$ correspond to some recursive call of \cref{algo:pareto_front}.
    Let
    \[
        T(\Pi) = \sum_{(I^i, R^i) \in \Pi} |I^i| \log \frac{n}{|I^i|}.
    \]
    Since $H(I)$ is the minimum over all respectful partitions, we have $T(\Pi) \ge n \cdot H(I)$,
    which implies $\Omega(T(\Pi)) \supseteq \Omega(n \cdot H(I))$.
    What remains is to show $\Runtime(\Apf, I) \in \Omega(n + T(\Pi)))$.
    Re-index the $(R^i, I^i)$ to be sorted by recursion depth.
    Then, since the recursion tree of \cref{algo:pareto_front} is a binary tree,
    the points in $I^i$ participate in at least $\log(i)$ levels of recursion,
    so \cref{algo:convex_hull} spends at least $\log(i)$ time per such point.
    Therefore,
    \[
        \Runtime(\Apf, I) \ge \sum_{(I^i, R^i) \in \Pi} |I^i| \log i.
    \]
    The \emph{log sum inequality} for sequences $a_1, \dots, a_k, b_1, \dots, b_k > 0$ with $a := \sum_{i=1}^{k} a_i$ and $b := \sum_{i=1}^{k} b_i$ states that
    \[
        \sum_{i=1}^{k} a_i \log \frac{a_i}{b_i} \ge a \log \frac{a}{b}.
    \]
    Multiply both sides by $n$ and rewrite this to
    \[
        \sum_{i=1}^{k} n \, a_i \log \frac{1}{b_i} \ge n\, a \log \frac{a}{b} + \sum_{i=1}^{k} n\, a_i \log \frac{1}{a_i}.
    \]
    Put $k = |\Pi|$, $a_i = \frac{|I^i|}{n}$ and $b_i = \frac{1}{i^2}$, with $a = 1$ and $1 \le b \le e$,
    then this is
    \[
        \sum_{i=1}^{k} |I^i| \log (i^2) \ge n \log \frac{1}{b} + \sum_{i=1}^{k} |I^i| \log \frac{n}{|I^i|} \ge -n + \sum_{i=1}^{k} |I^i| \log \frac{n}{|I^i|}.
    \]
    Therefore,
    \[
        2 \Runtime(\Apf, I) \ge 2 \sum_{(I^i, R^i) \in \Pi} |I^i| \log i \ge -n + \sum_{i=1}^{k} |I^i| \log \frac{n}{|I^i|} = -n + T(\Pi).
    \]
    Since  \cref{algo:convex_hull} always needs $\Omega(n)$ time to read the input,
    this shows $\Runtime(\Apf, I) \in \Omega(n + T(\Pi))$.
\end{proof}

\subsection{Convex Hull}

We now show part (ii) of:

\begin{lemma} \label{lemm:ch_algo_slow}
    Let $\Ach$ be \cref{algo:convex_hull}.
    Then $\Runtime(\Ach, I) \in \Omega(n(1+H(I)))$ for any input $I$.
\end{lemma}

\begin{proof}
    The proof is analogous to that of \cref{lemm:algo_slow}.
\end{proof}

\convexTHM*

\begin{proof}[Proof of \cref{thm:convex} part (ii) using \cref{lemm:ch_algo_slow}]
    The proof is analogous to that of \cref{thm:pareto} part (ii).
\end{proof}

\bibliographystyle{plainurl}
\bibliography{references}

\end{document}